\documentstyle[seceq,preprint,epsf]{jpsj}

\makeatletter
\def\lesssim{\mathrel{\mathpalette\vereq<}}

\def\vereq#1#2{\lower3pt\vbox{\baselineskip1.5pt \lineskip1.5pt
\ialign{$\m@th#1\hfill##\hfil$\crcr#2\crcr\sim\crcr}}}
\makeatother

\def\runtitle{
Perturbation Study of the Conductance through 
an Interacting Region Connected to Multi-Mode Leads}

\def\runauthor{Yoshihide {\sc Tanaka}, Akira {\sc Oguri} 
               and Hiroumi {\sc Ishii} }

\title
{
Perturbation Study of the Conductance through 
an Interacting Region Connected to Multi-Mode Leads  
}

\author{ Yoshihide {\sc Tanaka}, Akira  {\sc Oguri} and Hiroumi {\sc Ishii} }

\inst{
Department of Material Science, 
Osaka City University, Sumiyoshi-ku, Osaka 558-8585, Japan
}

\recdate{August 31, 2001}

\abst{
We study the effects of electron correlation on 
transport through an interacting region 
connected to multi-mode leads
based on the perturbation expansion with respect to 
the inter-electron interaction.
At zero temperature the conductance defined in the Kubo formalism
can be written in terms of a single-particle Green's function 
at the Fermi energy, and it can be mapped onto a transmission 
coefficient of the free quasiparticles described by an 
effective Hamiltonian. 
We apply this formulation to a two-dimensional Hubbard model
of finite size connected to two noninteracting leads.  
We calculate the conductance in the electron-hole symmetric case
using the order $U^2$ self-energy.
The conductance shows several maximums in the $U$ dependence 
in some parameter regions of $t_y/t_x$,
where $t_x$ ($t_y$) is the hopping matrix element 
in the $x$- ($y$-) directions.
This is caused by the resonance occurring in some of the subbands,
and is related with the $U$ dependence of 
the eigenvalues of the effective Hamiltonian. 
}

\kword
{
conductance, subband structure, 
electron correlation, 
perturbation expansion, 
Hubbard model, 
two-dimension 
}

\begin{document}
\sloppy
\maketitle

\section{Introduction}
Low-dimensional electron systems have been 
one of the current interests in the fields of 
the condensed matter physics and materials science. 
For instance, in some of the organic conductors, 
the electron correlation has been considered to be important 
to understand the physical properties.~\cite{ishiguro} 
The Kondo effect in quantum dots~\cite{NL,GR,Kawabata} has 
also been studied intensively from theoretical~\cite{MWL1-2,HDW2} and 
experimental~\cite{RalphBuhrman,Goldharber-Gordon,Kouwenhoven} 
sides.
When the average number of the electrons in a quantum dot is odd,
the perfect transmission due to the Kondo resonance situated 
at the Fermi energy occurs at low temperatures. 
Recently, artificial molecules which are realized 
by arranging two or more quantum dots 
have also been studied.~\cite{oosterkamp,tokura}
Theoretically, the crossover from the high-temperature Coulomb-blockade 
to low-temperature Fermi-liquid behaviors of the quantum dots  
has been studied  using advanced numerical methods 
such as the numerical renormalization 
group~\cite{Izumida1-3,Izumida4-5} and 
quantum Monte Carlo methods.~\cite{Sakai,ao,ao6}

In a previous work, one of the authors
has studied the conductance of small interacting systems
connected to two single-mode leads,~\cite{ao7,ao9}
and has calculated the conductance of a Hubbard chain of 
finite size $N$ using the order $U^2$ self-energy.  
The results obtained in the electron-hole symmetric case
depend strongly on whether $N$ is even or odd.
In the even cases, the conductance decreases 
with increasing $N$ showing a tendency 
toward a Mott-Hubbard insulator.
On the other hand, in the odd cases the perfect transmission  
due to the Kondo resonance occurs.\cite{ao9}

The purpose of the present work 
is to generalize the formulation to the multi-mode systems 
where the interacting system is connceted 
to noninteracting leads with a number of channels.
As in the single-mode case, 
at $T=0$ the contributions of the vertex corrections on
the dc conductance $g$ vanish.
Then the conductance is determined by 
the value of the single-particle Green's function 
at the Fermi energy $\omega=0$, 
and thus $g$ can be written in terms of the transmission coefficient 
of the free quasiparticles described by an effective Hamiltonian 
also in the multi-mode case.
We apply the formulation to a two-dimensional Hubbard model 
of finite size consisting of $N_{\rm C}=N \times M$ sites, 
where $N$ and $M$ are the size in the $x$- and $y$-direction,
respectively. Two noninteracting leads of $M$ channels 
are attached to this cluster along the $x$-direction.
This system may be considered as a model for 
two-dimensional materials such as 
an array of quantum dots and a carbon nanotube. 
For instance, for $M=2$ it may also be considered as a model 
for a ladder system of nanometer size.
Since there is no translational symmetry in the systems 
we are considering, the self-energy has $N_{\rm C} \times N_{\rm C}$ matrix elements. 
We calculate all these elements within the second order 
perturbation expansion in $U$ in 
the electron-hole symmetric case, 
and determine the parameters of the effective Hamiltonian up to 
terms of order $U^2$. 
The results of the conductance show maximums 
at finite $U$ for some values of $t_y/t_x$, where $t_x$ ($t_y$) is 
the nearest-neighbor transfer integral in the $x$-direction ($y$-direction).
This is caused by resonances occurring in some of the subbands. 
These behaviors can also be understood through 
the $U$ dependence of the eigenvalues 
of the effective Hamiltonian.
We note that the second-order perturbation theory has been 
applied by a number of groups 
to study transport through 
the single Anderson impurity\cite{HDW2,YMF,MiiMakoshi,TakagiSaso}
and systems consisting of a number of sites.\cite{PFA,KKNO} 
Also, the band calculations has been applied to obtain
the conductance of nanometer systems.~\cite{Lang} 
Our approach can, in principle, be extended to realistic systems 
by evaluating the order $U^2$ self-energy 
using the orbitals which are obtained with the band calculations.

In \S 2, we describe the outline of the general formulation and
introduce the effective Hamiltonian.
In \S 3, we apply the method to a two-dimensional Hubbard model,
and present the results of the conductance.
Summary is given in \S 4. 
In Appendix, we provide the derivation of the expressions of  
the conductance and the total charge displacement.

\section{Formulation}
\label{sec:Model}
In this section, 
we give the outline of a general formulation 
of the conductance through interacting systems 
connected to Fermi-liquid reservoirs with a number of channels.
Our formulation is applicable for various systems 
which have the time reversal symmetry, i.e., 
eqs.\ (\ref{eq:g_multi}) and (\ref{eq:Friedel_multi}) 
hold for a wide range of the systems 
described by the Hamiltonian eq.\ (\ref{eq:H}).
We provide the details of the derivations in Appendix,
and demonstrate the application 
to a Hubbard model in the next section.

\begin{figure}
\hspace{0.5cm}
\centerline{ \vbox{ \epsfxsize=150mm \epsfbox {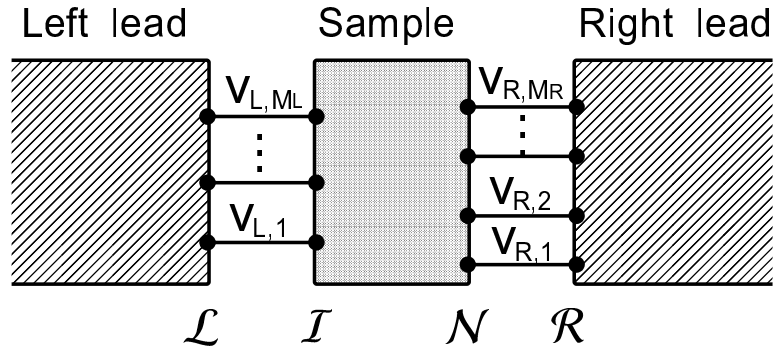}}
}
\vspace{-13.5cm}
\caption{  Schematic picture of the system.} 
\label{fig:multi}
\end{figure}

We start with a system which consists of three regions
as illustrated in Fig.\ \ref{fig:multi};
a finite interacting region (${\rm C}$) situated in the middle,
and two noninteracting reservoirs on the left (${\rm L}$) and right (${\rm R}$). 
The central region contains $N_{\rm C}$ resonant levels, 
and the inter-electron interaction $U_{j_4 j_3; j_2 j_1}$ is 
switched on in the central region.
Each of the two reservoirs is infinitely large and 
has a continuous energy spectrum.
The central region and two reservoirs are connected 
by $M_{\rm L}$ and $M_{\rm R}$ channels which are described by 
the mixing matrix elements $v_{{\rm L},m}^{\phantom{\dagger}}$ 
and $v_{{\rm R},m}^{\phantom{\dagger}}$, respectively.
The complete Hamiltonian is given by 
\begin{eqnarray}
{\cal H} \ 
    &=&  \ {\cal H}_{\rm L} + {\cal H}_{\rm R}  + {\cal H}_{\rm C}^0 + 
{\cal H}_{\rm C}^{\rm int}
            + {\cal H}_{\rm mix}     
\label{eq:H}            
\;, \\
{\cal H}_{\rm L} &=&  \sum_{ij\in {\rm L}} \sum_{\sigma} 
        \left(\,-t_{ij}^{\rm L} - \mu\, \delta_{ij} \,\right)
            c^{\dagger}_{i \sigma} c^{\phantom{\dagger}}_{j \sigma}      
             \;, 
\label{eq:H_L}           
\\ 
{\cal H}_{\rm R}  &=&  \sum_{ij\in {\rm R}} \sum_{\sigma} 
        \left(\,-t_{ij}^{\rm R} - \mu\, \delta_{ij} \,\right)
  c^{\dagger}_{i \sigma} c^{\phantom{\dagger}}_{j \sigma}        
      \;, 
\label{eq:H_R}      
\\      
  {\cal H}_{\rm C}^{0} &=&   \sum_{ij\in {\rm C}} \sum_{\sigma} 
        \left(\,-t_{ij}^{\rm C} - \mu\, \delta_{ij} \,\right)  
   c^{\dagger}_{i \sigma} c^{\phantom{\dagger}}_{j \sigma}        
  \;, \\   
    {\cal H}_{\rm C}^{\rm int} &=&     
{1 \over 2} \sum_{\{j\} \in {\rm C}}\sum_{\sigma \sigma'}    
    U_{j_4 j_3; j_2 j_1}\,  
 c^{\dagger}_{j_4 \sigma} c^{\dagger}_{j_3 \sigma'}    
 c^{\phantom{\dagger}}_{j_2 \sigma'} c^{\phantom{\dagger}}_{j_1 \sigma} 
  \;,  
\label{eq:H_int}  
\\
  {\cal H}_{\rm mix} &=& 
        - \sum_{m=1}^{M_{\rm L}} \sum_{\sigma} 
                  v_{{\rm L},m}^{\phantom{\dagger}} 
           \left(\, 
            c^{\dagger}_{{\cal I}_m \sigma}\, 
            c^{\phantom{\dagger}}_{{\cal L}_m \sigma}  
          + c^{\dagger}_{{\cal L}_m \sigma}\, 
            c^{\phantom{\dagger}}_{{\cal I}_m \sigma}
\, \right) 
\nonumber \\
& & - \sum_{m=1}^{M_{\rm R}} \sum_{\sigma}
          v_{{\rm R},m}^{\phantom{\dagger}}
  \left(\, 
           c^{\dagger}_{{\cal R}_m \sigma}\, 
           c^{\phantom{\dagger}}_{{\cal N}_m  \sigma} 
        +  c^{\dagger}_{{\cal N}_m\, \sigma} 
           c^{\phantom{\dagger}}_{{\cal R}_m  \sigma} 
\, \right) .
\label{eq:H_mix_multi}
\end{eqnarray}
Here $c^{\dagger}_{j \sigma}$ 
 ($c^{\phantom{\dagger}}_{j \sigma}$) creates (destroys) 
an electron with spin $\sigma$ at site $j$,  
and $\mu$ is the chemical potential.
$t_{ij}^{\rm L}$, $t_{ij}^{\rm R}$, and $t_{ij}^{\rm C}$ are 
the intra-region hopping matrix elements
in each of the regions ${\rm L}$, ${\rm R}$, and ${\rm C}$, respectively. 
The labels $1$, $2$, $\ldots$, $N_{\rm C}$ are assigned to 
the sites in the central region.
In eq.\ (\ref{eq:H_mix_multi}), 
 ${\cal I}_m$ (${\cal L}_m$) is the label 
assigned to the $m$-th site 
at the sample side (lead side) of the interface on the left. 
${\cal N}_m$ (${\cal R}_m$) is the label assigned to the $m$-th site 
at the sample side (lead side) of the interface on the right.
Note that the number of the channels is less than 
the number of the interacting sites, i.e,
$M_{\rm L} \leq N_{\rm C}$ and $M_{\rm R}\leq N_{\rm C}$. 
For instance, in the case of $M_{\rm L}=M_{\rm R}=N_{\rm C}$, 
all the sites in the central region are connected to both of the reservoirs,
i.e., ${\cal I}_m = {\cal N}_m$ for $m=1,\,2,\,\ldots,\, N_{\rm C}$. 
We assume that all the hopping matrix elements are real,
and the interaction has the time reversal symmetry:  
$U_{4 3; 2 1}$ is real 
and $U_{4 3; 2 1}=U_{3 4; 1 2}=U_{1 2; 3 4 }=U_{4 2; 3 1}=U_{1 3; 2 4}$. 
We will be using units $\hbar=1$ unless otherwise noted.

The single-particle Green's function is defined by 
\begin{equation} 
G_{jj'}({\rm i}\varepsilon_n) 
 =  
-    \int_0^{\beta} \! {\rm d}\tau 
   \left \langle  T_{\tau} \,  
   c^{\phantom{\dagger}}_{j \sigma} (\tau) \, c^{\dagger}_{j' \sigma} (0)     
                   \right \rangle  \, {\rm e}^{{\rm i}\, \varepsilon_n \tau} .
  \label{eq:G_Matsubara}                  
\end{equation} 
Here $\beta= 1/T$, $\varepsilon_n = (2n+1)\pi/\beta$, 
$c_{j \sigma}(\tau) = 
{\rm e}^{\tau  {\cal H}} c_{j \sigma} {\rm e}^{- \tau  {\cal H}}$,
and $\langle \cdots \rangle$ denotes the thermal average 
$\mbox{Tr} \left[ \, {\rm e}^{-\beta  {\cal H} }\, {\cdots}
\,\right]/\mbox{Tr} \, {\rm e}^{-\beta  {\cal H} }$.
The spin index has been omitted from the left-hand side 
of eq.\ (\ref{eq:G_Matsubara}) 
assuming the expectation value to be 
independent of whether spin is up or down.
Since the interaction is switched on only 
for the electrons in the central region,
the Dyson equation can be written as
\begin{equation} 
  G_{ij}(z)    =   G^0_{ij}(z) 
    + \sum_{i'j' \in {\rm C}}\,G^0_{ii'}(z)\,  \Sigma_{i'j'}(z)  
   \, G_{j'j}(z) \;.    
  \label{eq:Dyson}   
\end{equation}   
Here $G^0_{ij}(z)$ is the unperturbed Green's function 
corresponding to the noninteracting 
Hamiltonian ${\cal H}^0  \equiv   {\cal H}_{\rm L} + {\cal H}_{\rm R}  
   + {\cal H}_{\rm C}^0 + {\cal H}_{\rm mix}$.
The summations with respect to $i'$ and $j'$ run over
the sites in the central region, 
and $\Sigma_{i'j'}(z)$ is the self-energy 
due to the interaction ${\cal H}_{\rm C}^{\rm int}$. 
Because of the time reversal symmetry of ${\cal H}$, 
these functions are symmetric against the interchange
of the site indices: 
 $G_{ij}(z) =  G_{ji}(z)$ and $\Sigma_{ij}(z) = \Sigma_{ji}(z)$. 
Note that at $T=0$ the single-particle excitation 
at the Fermi energy $z= {\rm i}0^+$ does not decay, i.e.,
 $\mbox{Im}\,\Sigma_{ij}^+ (0)=0$.\cite{LangerAmbegaokar}  
In what follows, we will treat $z$ as a complex variable,
and use the symbol $+$ ($-$) in the superscript 
as a label for the retarded (advanced) function:  
$\Sigma_{ij}^{\pm}(\varepsilon) \equiv 
\Sigma_{ij}(\varepsilon \pm {\rm i}0^+)$.

As in the single-mode case, 
the dc conductance at $T=0$ can be expressed 
in terms of the single-particle Green's function 
at $\omega=0$ [see Appendix \ref{sec:CONDUCTANCE}];   
\begin{equation}
 g =  {2 e^2 \over h} \, 
               \mbox{Tr} \left[\,4\,
               \mbox{\boldmath $\Gamma$}_{\rm R} (0) \, 
               \mbox{\boldmath $G$}_{{\cal N}{\cal I}}^{+}(0)\, 
               \mbox{\boldmath $\Gamma$}_{\rm L}(0) \,  
               \mbox{\boldmath $G$}_{{\cal I}{\cal N}}^{-}(0)  
               \,\right] . 
\label{eq:g_multi}
\end{equation}
Here $\mbox{\boldmath $\Gamma$}_{\rm L}(\omega)$ 
and $\mbox{\boldmath $\Gamma$}_{\rm R}(\omega)$ 
are $M_{\rm L} \times M_{\rm L}$ and  $M_{\rm R} \times M_{\rm R}$ matrices, 
respectively. 
These two matrices are caused by the coupling with the left (${\rm L}$) 
and right (${\rm R}$) leads,  
and the elements are given by $\Gamma_{\alpha;mm'} (\omega) = 
- \mbox{Im} \left[ v_{\alpha,m}\,  
F_{\alpha,mm'}^{+}(\omega) \,v_{\alpha,m'} \right]$
with $F_{\alpha,mm'}(z)$ being the Green's functions 
at the interface of the isolated lead ($\alpha ={\rm L},{\rm R}$).
In eq.\ (\ref{eq:g_multi}), 
$\mbox{\boldmath $G$}_{{\cal N}{\cal I}}^+$ 
and $\mbox{\boldmath $G$}_{{\cal I}{\cal N}}^-$ 
are  $M_{\rm R} \times M_{\rm L}$ and $M_{\rm L} \times M_{\rm R}$ matrices 
the elements of which are given by 
$G_{{\cal N}_{l}{\cal I}_{m}}^+$ 
and  $G_{{\cal I}_{l}{\cal N}_{m}}^-$, respectively.

Another quantity which can be related to the scattering coefficients 
is the displacement of the total charge defined by
\cite{LangerAmbegaokar,Langreth} 
\begin{eqnarray}
\Delta N_{\rm tot}
\ &=&  \ 
\sum_{i\in {\rm C}} \sum_{\sigma} 
 \langle c^{\dagger}_{i \sigma} c^{\phantom{\dagger}}_{i \sigma}
\rangle 
\nonumber\\
& &
+ \sum_{i\in {\rm L}} \sum_{\sigma} 
  \left[\, 
  \langle c^{\dagger}_{i \sigma} c^{\phantom{\dagger}}_{i \sigma} \rangle
  - 
  \langle c^{\dagger}_{i \sigma} 
  c^{\phantom{\dagger}}_{i \sigma} \rangle_{\rm L}^{\phantom{0}}
    \,\right] 
\nonumber\\
& &
+ \sum_{i\in {\rm R}} \sum_{\sigma} 
  \left[\, 
  \langle c^{\dagger}_{i \sigma} c^{\phantom{\dagger}}_{i \sigma} \rangle
  - 
  \langle c^{\dagger}_{i \sigma} 
  c^{\phantom{\dagger}}_{i \sigma} \rangle_{\rm R}^{\phantom{0}}
    \,\right]. 
\label{eq:dn_def}
\end{eqnarray}
Here $\,\langle \cdots \rangle_{\rm L}^{\phantom{0}}$  and 
$\,\langle \cdots \rangle_{\rm R}^{\phantom{0}}$  
denote the ground-state average of isolated leads
described by ${\cal H}_{\rm L}$ and ${\cal H}_{\rm R}$, respectively.
At $T=0$, following the derivation 
of the Friedel sum rule 
by Langer and Ambegaokar,\cite{LangerAmbegaokar} 
$\Delta N_{\rm tot}$ can be written 
in terms of 
a $(M_{\rm L} +M_{\rm R}) \times (M_{\rm L} +M_{\rm R})$ matrix $\mbox{\boldmath $S$}$, 
\begin{eqnarray}
& &\Delta N_{\rm tot}
= 
{1 \over \pi {\rm i}}\, 
\log [\, \det 
 \mbox{\boldmath $S$}
\,] \;,
\label{eq:Friedel_multi}
\\
\nonumber
\\
& &\mbox{\boldmath $S$}
\, = \,  
\mbox{\boldmath $1$}
- \, 2 {\rm i} 
  \left [ 
 \matrix { \mbox{\boldmath $\Gamma$}_{\rm L}(0)  & \mbox{\boldmath $0$}  \cr
           \mbox{\boldmath $0$}  & \mbox{\boldmath $\Gamma$}_{\rm R}(0)  \cr  }
  \right ]  
  \left [ 
 \matrix {   \mbox{\boldmath $G$}_{{\cal I}{\cal I}}^{+}(0)  
           & \mbox{\boldmath $G$}_{{\cal I}{\cal N}}^{+}(0)  \cr 
             \mbox{\boldmath $G$}_{{\cal N}{\cal I}}^{+}(0)  
           & \mbox{\boldmath $G$}_{{\cal N}{\cal N}}^{+}(0)  \cr  }
  \right ]  
     .   
\nonumber     
\\ 
\label{eq:S_multi}
\end{eqnarray}
Here $\mbox{\boldmath $G$}_{{\cal I}{\cal I}}^+$  
and $\mbox{\boldmath $G$}_{{\cal N}{\cal N}}^+$ 
are $M_{\rm L} \times M_{\rm L}$ 
and $M_{\rm R} \times M_{\rm R}$ matrices,
and the elements are given by $G_{{\cal I}_{l}{\cal I}_{m}}^+$ 
and $G_{{\cal N}_{l}{\cal N}_{m}}^+$, respectively. 
The outline of the Friedel sum rule in the single-mode case 
is given in Appendix \ref{sec:Friedel}.
The extension to multi-mode case eq.\ (\ref{eq:Friedel_multi}) 
is straightforward because in the case of the Friedel sum rule 
we do not have to take into account the contributions vertex corrections. 
For convenience, in eq.\ (\ref{eq:S_multi}) we have assumed that 
there is no common element in the sets ${\cal I}$ and ${\cal N}$. 
This is not an important assumption, and 
the expression without the assumption can be obtained in the similar way. 
Note that eq.\ (\ref{eq:Friedel_multi}) 
is also written in terms of the Green's function 
at $T=0$ and $\omega=0$.
Thus, due to the property $\mbox{Im}\,\Sigma_{ij}^+ (0)=0$, 
the values of $g$ and $\Delta N_{\rm tot}$ at $T=0$ can be 
expressed in terms of the transmission and reflection coefficients 
defined with respect to a one-particle Hamiltonian\cite{ao9} 
\begin{equation} 
 \widetilde{\cal H}_{\rm qp}  =   {\cal H}_{\rm L} + {\cal H}_{\rm R}  
   + {\cal H}_{\rm C}^{\rm eff} + {\cal H}_{\rm mix}\;.
\label{eq:H_eff}
\end{equation} 
Here ${\cal H}_{\rm C}^{\rm eff} 
 =  \sum_{ij\in {\rm C},\sigma} 
   \left(\,-\widetilde{t}_{ij}^{\rm C} - \mu\, \delta_{ij} \,\right)  
   c^{\dagger}_{i \sigma} c^{\phantom{\dagger}}_{j \sigma}$ 
with $-\widetilde{t}_{ij}^{\rm C} =  -t_{ij}^{\rm C} + \mbox{Re}\, \Sigma_{ij}^+ (0)$. 
This effective Hamiltonian describes 
free quasiparticles of the local Fermi-liquid.
As shown in the next section, 
the eigenvalues of the partial Hamiltonian ${\cal H}_{\rm C}^{\rm eff}$ have
important information about the ground-state properties.

\section{Application to a Two-dimensional Hubbard Model}
\label{sec:Hubbard2D}
In this section, 
we apply the formulation to a two-dimensional Hubbard model
connected to reservoirs and calculate the conductance 
using the order $U^2$ self-energy 
in the electron-hole symmetric case.

\subsection{Model}

\begin{figure}
\vspace{-5cm}
\hspace{1cm}
\centerline{ \vbox{ \epsfxsize=150mm \epsfbox {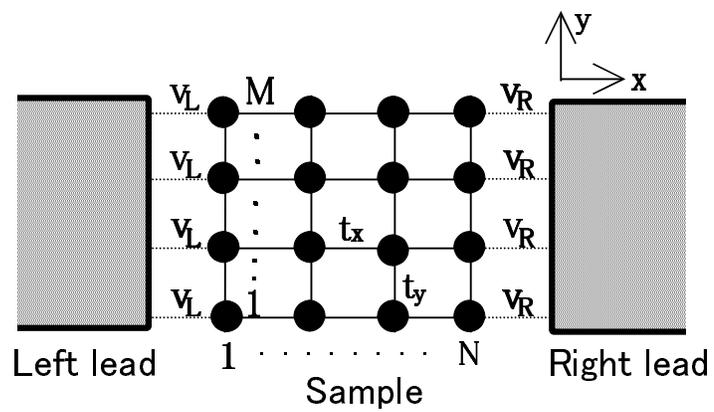}}
}
\vspace{-8cm}
\caption{Schematic picture of a 2D Hubbard model:
($\bullet$) interacting sites.} 
\label{fig:2dHubbard}
\end{figure}

\begin{figure}
\vspace{-5cm}
\hspace{3.5cm}
\centerline{ \vbox{ \epsfxsize=150mm \epsfbox {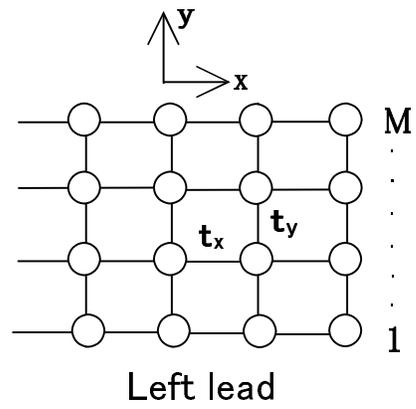}}
}
\vspace{-8.5cm}
\caption{ Schematic picture of the type I lead.} 
\label{fig:2dHubbard_2}
\end{figure}

The schematic picture of the model 
is illustrated in Fig.\ \ref{fig:2dHubbard}.   
The system size in the $y$-direction, $M$, is finite.  
In the $x$-direction, 
the central region consists of $N$ columns, and
two noninteracting leads are connected at $x=1$ and $x=N$.
Thus, the total number of the interacting sites is $N_{\rm C} = N\times M$.
The parameters of the Hamiltonian eq.\ (\ref{eq:H}) are 
specified as follows.  
The mixing matrix element is assumed to be uniform in the $y$-direction:
$v_{{\rm L},m}^{\phantom{0}} =v_{\rm L}^{\phantom{0}}$ and 
$v_{{\rm R},m}^{\phantom{0}} = v_{\rm R}^{\phantom{0}}$ for $m=1,2,\ldots,M$.
We assume that the off diagonal part of 
$t_{ij}^{\rm C}$ describes the nearest-neighbor hopping; 
$t_x$ and $t_y$ in the $x$- and $y$-directions, respectively.
Along the $y$-direction, we use the periodic boundary condition. 
Furthermore,
we assume $U_{j_4 j_3; j_2 j_1}$ to be an onsite repulsion $U$, 
and  concentrate on the electron-hole symmetric case
taking the parameters to be $\mu =0$ and $\epsilon_d + U/2 =0$, 
where $\epsilon_d$ is the onsite energy 
of the interacting sites $-t_{ii}^{\rm C}=\epsilon_d$.
Then the Dyson equation can be written in a $N_{\rm C}\times N_{\rm C}$ matrix form:
$
\left\{ \widehat{\mbox{\boldmath ${\cal G}$}}(z) \right\}^{-1} =
\left\{ \widehat{\mbox{\boldmath ${\cal G}$}}^0(z) \right\}^{-1} 
 - \widehat{\mbox{\boldmath $\Sigma$}}(z)$ with 
\begin{eqnarray}
\left\{ \widehat{\mbox{\boldmath ${\cal G}$}}^0(z) \right\}^{-1} 
&=&
z \, \widehat{\mbox{\boldmath $1$}}
 - \widehat{\mbox{\boldmath ${\cal H}$}}_{\rm C}^0 
- \widehat{\mbox{\boldmath ${\cal V}$}}_{\rm mix}(z) \;.
\label{eq:81}
\end{eqnarray}
Here $\widehat{\mbox{\boldmath $1$}}$ is 
the $N_{\rm C} \times N_{\rm C}$ unit matrix, and the matrices can be 
written in the partitioned forms; 
\begin{eqnarray}
\widehat{\mbox{\boldmath ${\cal H}$}}_{\rm C}^0 &=&
\left[ \matrix{
 \mbox{\boldmath $h$}_y^0& -t_x \mbox{\boldmath $1$} 
 &   & \mbox{\boldmath $0$} \cr
-t_x \mbox{\boldmath $1$} & \ddots          & \ddots &   \cr
          & \ddots          & \ddots & -t_x \mbox{\boldmath $1$} \cr
\mbox{\boldmath $0$} &   & -t_x \mbox{\boldmath $1$} 
& \mbox{\boldmath $h$}_y^0 \cr
 } \right],
\label{eq:85} 
\\
\widehat{\mbox{\boldmath ${\cal V}$}}_{\rm mix}(z) &=&
\left[
\matrix{
v_{\rm L}^2 \mbox{\boldmath $F$}_{\rm L}(z) 
& & & & \cr
 & & & &  \cr
 & & \mbox{\boldmath $0$} & &  \cr
 & & & &  \cr
 & & & & 
 v_{\rm R}^2 \mbox{\boldmath $F$}_{\rm R}(z)
 \cr
 } \right],
\label{eq:87}
\\
\widehat{\mbox{\boldmath $\Sigma$}}(z) &=& 
\left[
\matrix{
\mbox{\boldmath $\Sigma$}_{11}(z) & 
\mbox{\boldmath $\Sigma$}_{12}(z) & \ldots & 
\mbox{\boldmath $\Sigma$}_{1N}(z) \cr
\mbox{\boldmath $\Sigma$}_{21}(z) & \ddots                  
& \ddots & \vdots                  \cr
\vdots        & \ddots           & \ddots & \vdots                  \cr
\mbox{\boldmath $\Sigma$}_{N1}(z) & \ldots 
  & \ldots & \mbox{\boldmath $\Sigma$}_{NN}(z) \cr
 }  \right],
\label{eq:86}
\end{eqnarray}
where 
$\mbox{\boldmath $1$}$ is the $M\times M$ unit matrix, and
$\mbox{\boldmath $h$}_y^0$ is the $M\times M$ hopping matrix 
in the $y$-direction;  
\begin{equation}
\mbox{\boldmath $h$}_y^0 \,=
\left [
\matrix{
0    & -t_y    &        &        &         & -t_y \cr
-t_y & 0       & -t_y   &        & \mbox{\boldmath $0$} &      \cr
     & -t_y    & 0      & \ddots &         &      \cr
     &         & \ddots & \ddots & \ddots  &      \cr
     & \mbox{\boldmath $0$} &        & \ddots & \ddots  & -t_y \cr
-t_y &         &        &        & -t_y    & 0    \cr
 }
 \right].
\label{eq:59}
\end{equation}
We note that the Hartree-Fock term of the self-energy 
is already included in the unperturbed Green's function   
 $\widehat{\mbox{\boldmath ${\cal G}$}}^0(z)$ defined by eq.\ (\ref{eq:81}). 
Therefore, 
$\widehat{\mbox{\boldmath $\Sigma$}}(z)$ is 
the self-energy correction beyond the mean-field theory,
which is described by the many-body perturbation theory with respect to 
$
{\cal H}_{\rm C}^{\rm int} = U \sum_{i=1}^{N_{\rm C}} \left[\, 
                 n_{i \uparrow}\, n_{i \downarrow}
-   (   n_{i \uparrow}
                   + n_{i \downarrow} )/2
            \,\right]
$ where $n_{i \sigma} = 
 c^{\dagger}_{i \sigma} c^{\phantom{\dagger}}_{i \sigma}$. 
We note that the mixing with the noninteracting leads is included
in the unperturbed Hamiltonian, and 
thus $\widehat{\mbox{\boldmath $\Sigma$}}(z)$ depends 
on $v_{\rm L}$ and $v_{\rm R}$ through 
$\widehat{\mbox{\boldmath ${\cal G}$}}^0(z)$. 
In eq.\ (\ref{eq:86}) the partitioned element 
$\mbox{\boldmath $\Sigma$}_{ll'}(z)$ is 
a $M \times M$ matrix the $(m,m')$ element of which    
corresponds to the self-energy correction between the sites located 
at $\mbox{\boldmath $r$}=(l, m)$ and $\mbox{\boldmath $r$}'=(l', m')$, 
where $l$ and $m$ correspond to the $x$ and $y$ coordinates, respectively. 
The local Green's functions at the interfaces of the isolated leads, 
$\mbox{\boldmath $F$}_{\rm L}(z)$ and $\mbox{\boldmath $F$}_{\rm R}(z)$ 
in eq.\ (\ref{eq:87}), depend on the excitation spectrum of the leads, 
i.e., ${\cal H}_{\rm L}$ and ${\cal H}_{\rm R}$. 
We concentrate on the case 
$\mbox{\boldmath $F$}_{\rm L}=\mbox{\boldmath $F$}_{\rm R}$  
 ($\equiv \mbox{\boldmath $F$}$) 
assuming the same excitation spectrum for the left and right leads.
Then the matrices of the level width 
are given by 
$\mbox{\boldmath $\Gamma$}_{\rm L}(\omega)
= -v_{\rm L}^2\, \mbox{Im}\, \mbox{\boldmath $F$}(\omega+{\rm i} 0^+)$
and
$\mbox{\boldmath $\Gamma$}_{\rm R}(\omega)
= -v_{\rm R}^2 \,\mbox{Im}\,\mbox{\boldmath $F$}(\omega+{\rm i} 0^+)$. 
For the function $\mbox{\boldmath $F$}(z)$, we consider two types models: 
I) semi-infinite tight-binding leads, and 
II) leads of a constant density of states. 
A schematic picture of the type I lead
is illustrated in Fig.\ \ref{fig:2dHubbard_2}:   
the hopping matrix element is given by 
the nearest neighbor one, $t_x$ and $t_y$, as that in the central region. 
Therefore the Green's function for the type I lead 
$\mbox{\boldmath $F$}^{\rm I}$ satisfies a $M\times M$ matrix equation 
$\mbox{\boldmath $F$}^{\rm I}(z) 
= \left[\, z  
\mbox{\boldmath $1$} -  
\mbox{\boldmath $h$}_y^0 -
t_x^2\, \mbox{\boldmath $F$}^{\rm I}(z) \,\right]^{-1}$. 
In the case of the type II leads, 
we assume that the local density of states
at the interfaces $\rho$ is a constant 
and the bandwidth is infinity. 
Then the corresponding retarded Green's function becomes 
pure imaginary independent of the frequency $\omega$. 
Specifically, we consider a simple diagonal matrix of the form 
$\mbox{\boldmath $F$}^{\rm II}(\omega+{\rm i} 0^+) =
-{\rm i}\,\pi \rho \,\mbox{\boldmath $1$}$.
Thus for the type II leads, the effects of the mixing 
are parametrized by  
the constant $\Gamma_{\alpha} = \pi \rho \, v^2_{\alpha}$ 
for $\alpha ={\rm L}$ and ${\rm R}$.

The subband structure of the system  
is determined by the eigenstates of eq.\ (\ref{eq:59}): 
$\mbox{\boldmath $h$}_y^0 \, \mbox{\boldmath $\chi$}_n = 
\epsilon_n \, \mbox{\boldmath $\chi$}_n$ 
for $n=1,2,\ldots, M$.
Due to the translational symmetry in the $y$-direction, 
the self-energy $\mbox{\boldmath $\Sigma$}_{ll'}(z)$  
and the $M \times M$ matrix 
Green's function $\mbox{\boldmath $G$}_{ll'}(z)$, 
which are the $(l,l')$ partitioned 
element of $\widehat{\mbox{\boldmath ${\cal G}$}}(z)$,
can be diagonalized using the eigenstates;
\begin{eqnarray}
\mbox{\boldmath $\Sigma$}_{ll'}(z) 
 &=& \sum_{n=1}^M 
\mbox{\boldmath $\chi$}_n \, 
\Sigma_{ll'}^{(n)}(z) \,
\mbox{\boldmath $\chi$}_n^{\dagger}
\;,
\label{eq:self_mode}
\\
\mbox{\boldmath $G$}_{ll'}(z) 
 &=& \sum_{n=1}^M 
\mbox{\boldmath $\chi$}_n \, 
G_{ll'}^{(n)}(z) \,
\mbox{\boldmath $\chi$}_n^{\dagger}
\;.
\label{eq:Green_mode}
\end{eqnarray}
Therefore, 
the conductance
eq.\ (\ref{eq:g_multi}) 
can be decomposed into the sum of the contributions of the subbands: 
\begin{equation}
 g \, = \,  {2 e^2 \over h} \, 
             \sum_{n=1}^M 
               4\,
               \Gamma_{\rm R}^{(n)} (0) \, 
               G_{N1}^{(n)+}(0)\, 
               \Gamma_{\rm L}^{(n)}(0) \,  
               G_{1N}^{(n)-}(0)  
               \;. 
\label{eq:g_multi_mode}
\end{equation}
Here $\Gamma_{\alpha}^{(n)}(0)$ 
is defined by   
$\mbox{\boldmath $\Gamma$}_{\alpha}(z) 
 = \sum_{n=1}^M 
\mbox{\boldmath $\chi$}_n \, 
\Gamma_{\alpha}^{(n)}(0) \,
\mbox{\boldmath $\chi$}_n^{\dagger}
$ for $\alpha={\rm L},{\rm R}$. 
Similarly, 
the Friedel sum rule eq.\ (\ref{eq:Friedel_multi}) can be rewritten as
\begin{eqnarray}
& &\Delta N_{\rm tot}
= {2 \over \pi}
\sum_{n=1}^{M} 
{1 \over 2 {\rm i}} \,\log \left[
\det  \mbox{\boldmath $S$}^{(n)} \right]
\;,
\label{eq:Friedel_mode}
\\
\nonumber
\\
& &\mbox{\boldmath $S$}^{(n)}
 = 
 \left [ \,
 \matrix { 1  & 0  \cr
            0  & 1  \cr  }
 \, \right ]  
\nonumber \\
& & \qquad 
- \, 2 {\rm i} 
  \left [ 
 \matrix { \Gamma_{\rm L}^{(n)}(0)  & 0  \cr
            0  & \Gamma_{\rm R}^{(n)}(0)  \cr  }
  \right ]  
  \left [
 \matrix {   G_{11}^{(n)+}(0)  
           & G_{1N}^{(n)+}(0)  \cr 
             G_{N1}^{(n)+}(0)  
           & G_{NN}^{(n)+}(0)  \cr  }
 \right ] .   
\nonumber \\
\label{eq:S_mode}
\end{eqnarray}
Here $\delta^{(n)} \equiv 
1/(2{\rm i})\, \log 
\left[ \det\mbox{\boldmath $S$}^{(n)} \right]$  
corresponds to the phase shift of the $n$-th subband,  
and the charge displacement in each subband is given by
$\Delta N_{\rm tot}^{(n)} = 2\, \delta^{(n)}/\pi$.  
As mentioned, at $T=0$ the conductance is determined by the value 
of the Green's function at $\omega=0$, and
 $\mbox{Im}\,\widehat{\mbox{\boldmath $\Sigma$}}^+(0) 
= \widehat{\mbox{\boldmath $0$}}$ due to the Fermi-liquid property. 
Thus the effective Hamiltonian  
 $\widehat{\mbox{\boldmath ${\cal H}$}}_{\rm C}^{\rm eff} 
=\widehat{\mbox{\boldmath ${\cal H}$}}_{\rm C}^0 
+ \mbox{Re}\, \widehat{\mbox{\boldmath $\Sigma$}}^+(0)$,
which was introduced in the previous section,
can be used to calculate the conductance 
and the displacement of the total charge.\cite{ao9}
We now consider 
the $M\times M$ matrix $\mbox{\boldmath ${\cal H}$}_{{\rm C};ll'}^{\rm eff}$, 
which is the partitioned element of 
$\widehat{\mbox{\boldmath ${\cal H}$}}_{\rm C}^{\rm eff}$. 
The matrix $\mbox{\boldmath ${\cal H}$}_{{\rm C};ll'}^{\rm eff}$
can also be diagonalized as
\begin{eqnarray}
\mbox{\boldmath ${\cal H}$}_{{\rm C};ll'}^{\rm eff}
\, &=& \, - \sum_{n=1}^M 
\mbox{\boldmath $\chi$}_n \, 
\widetilde{t}^{(n)}_{ll'} 
\,
\mbox{\boldmath $\chi$}_n^{\dagger}
\;, 
\label{eq:H_eff_part}
\\
-\widetilde{t}^{(n)}_{ll'} 
\, &=& \, 
 \epsilon_n \, \delta_{ll'} \,
-t_x  \left[\,\delta_{l,l'+1} \, + \,  \delta_{l+1,l'} \,\right]
+ \mbox{Re}\,\Sigma_{ll'}^{(n)+}(0) 
\;,
\nonumber \\
\label{eq:25}
\end{eqnarray}
where $1\leq l, l' \leq N$. 
Thus each of the modes can be
mapped onto a tight-biding model in one-dimension 
with the renormalized hopping matrix element $\widetilde{t}^{(n)}_{ll'}$. 
As it will be seen later,
the behavior of eigenvalues of the $N \times N$ effective Hamiltonian 
defined by $\widetilde{\mbox{\boldmath ${\cal H}$}}_{\rm C}^{(n)} 
= \{ 
-\widetilde{t}^{(n)}_{ll'} 
\}$ is related to the transport properties.

\begin{figure}
\vspace{-8cm}
\hspace{0.5cm}
\centerline{ \vbox{ \epsfxsize=150mm \epsfbox {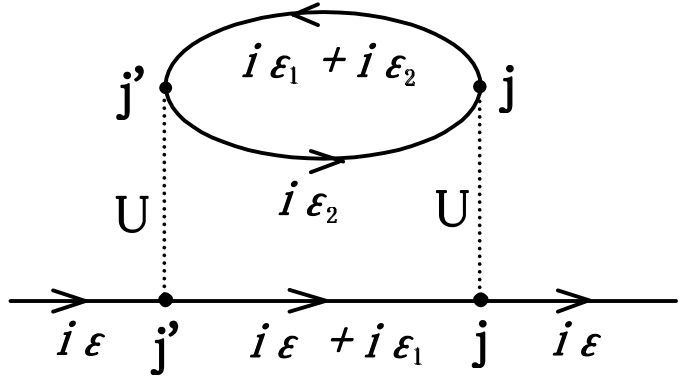}}
}
\vspace{-8cm}
\caption{The order $U^2$ self-energy $\Sigma_{jj'}({\rm i}\varepsilon)$.}
\label{fig:diagramSelf}
\end{figure}

We calculate the conductance using the perturbation expansion 
with respect to $U$.
In the electron-hole symmetric case, the order $U^2$ self-energy 
can be described by the diagram shown in Fig.\ \ref{fig:diagramSelf}; 
\begin{eqnarray}
 & &\Sigma_{jj'}^+(0)  \,  = 
\nonumber \\
& & \quad    
     - U^2   \int_{-\infty}^{\infty} \! \int_{-\infty}^{\infty} 
         \frac{{\rm d}\varepsilon\, {\rm d}\varepsilon'}{(2\pi)^2}
       \, G^{0}_{jj'}({\rm i} \varepsilon)  \,  
          G^{0}_{jj'}({\rm i} \varepsilon') \, 
          G^{0}_{j'j}({\rm i} \varepsilon + {\rm i} \varepsilon') 
\;,
\nonumber \\
 \label{eq:Self_2v} 
\end{eqnarray}
where $1\leq j, j' \leq N_{\rm C}$. 
The explicit form of the unperturbed 
Green's function $G^{0}_{jj'}({\rm i} \varepsilon)$  
can be obtained by taking the inverse of eq.\ (\ref{eq:81}).
Since ${\cal H}_{\rm mix}$ is included in the unperturbed part,
 $G^{0}_{jj'}$ and $\Sigma_{jj'}^+$ 
depend on the mixing matrix elements 
$v_{\rm L}^{\phantom{0}}$ and $v_{\rm R}^{\phantom{0}}$. 
Note that the retarded function at $\omega=0$ 
can be obtained from the Matsubara function, i.e.,   
$\Sigma_{jj'}^+(0) = \Sigma_{jj'}({\rm i}\varepsilon)|_{\varepsilon \to 0^+}$.
The imaginary part of eq.\ (\ref{eq:Self_2v}) vanishes, i.e., 
 $\mbox{Im} \Sigma_{jj'}^+(0)=0$ owing to the Fermi-liquid property. 
Furthermore, due to the electron-hole symmetry, 
 $\mbox{Re}\, \Sigma_{jj'}^+(0) = 0$ if $j$ and $j'$ belong 
 to the same sublattice. 
We calculate all the nonzero elements of $\mbox{Re}\, \Sigma_{jj'}^+(0)$ 
carrying out the integrations in eq.\ (\ref{eq:Self_2v}) numerically. 
Then the full Green's function $\mbox{\boldmath $G$}_{N1}^{+}(0)$ 
is evaluated substituting the order $U^2$ self-energy 
into the Dyson equation. 
In what follows we will assume the inversion 
symmetry $v_{\rm L}^{\phantom{0}} =v_{\rm R}^{\phantom{0}}$ ($\equiv v$),
and take the parameter to be $v/t_x= 0.9$ in the case of the type I leads. 
We have actually done some calculations also for $v/t_x= 0.7$,  
but the results are similar to those for $v/t_x= 0.9$ qualitatively.
Correspondingly, in the case of the type II leads  
we set $\Gamma_{\rm L} = \Gamma_{\rm R}$ ( $\equiv\ \Gamma$ ) 
and take the value to be $\Gamma/t_x= 0.75$.
We will mainly discuss the results obtained 
for the leads of the width $M=4$.

\subsection{Noninteracting systems}

\begin{figure}
\vspace{-1cm}
\hspace{1cm}
\centerline{ \vbox{ \epsfxsize=150mm \epsfbox {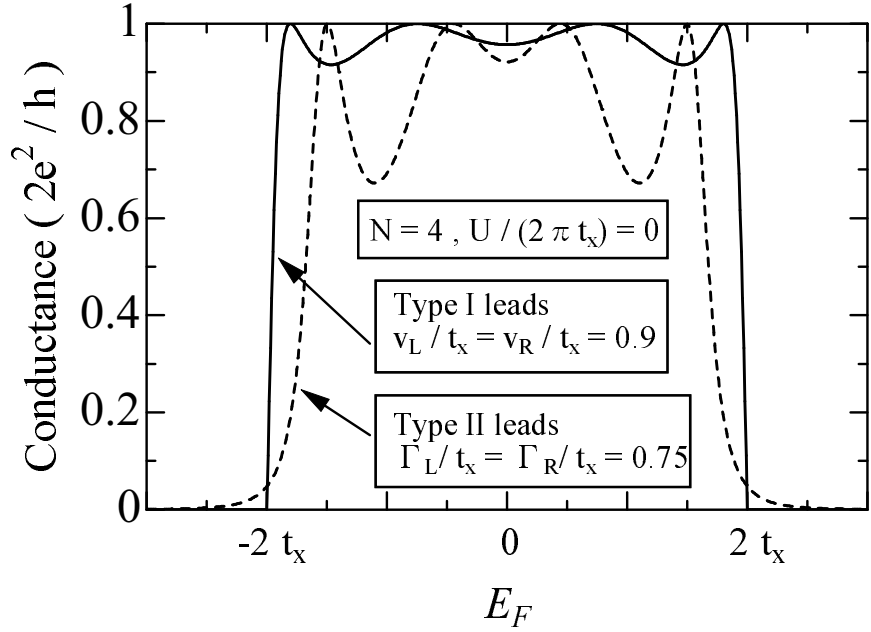}}
}
\vspace{-13cm}
\caption{Conductance vs.\ $E_F$ 
for noninteracting electrons on the one-dimensional chain 
of the size $N=4$ 
connected to the leads; (solid line) type I leads 
and (dashed line) type II leads.}
\label{fig:40}
\end{figure}

We now discuss some properties in the noninteracting case
in order to show the situations we are considering clear.
Fig.\ \ref{fig:40} shows 
the conductance $g_{\rm 1d}^0(E_F)$ 
for noninteracting electrons in the one-dimensional systems 
connected to the leads of the type I (solid line) and II (dashed line).
In this figure, $E_F$ is the Fermi-energy, and
the size of the central region is taken to be $N=4$.
The coupling with the leads are 
taken to be $v/t_x=0.9$ for the type I leads, 
and $\Gamma/t_x= 0.75$ for the type II leads.
There are $N$ ($=4$) peaks of the resonance,
and the peaks become sharp if the mixing matrix element 
 $v/t_x$ or $\Gamma/t_x$ decreases. 
In two dimensional lattice 
with the periodic boundary condition in the $y$-direction,
the conductance for noninteracting electrons 
is given by  $g_{\rm 2d}^0 = \sum_{n=1}^M \, g_{\rm 1d}^0(\epsilon_n)$
in the electron-hole symmetric case.
For instance, in the case of $M=4$, 
the eigenvalues $\epsilon_n$ are given 
by $\epsilon_1= -\epsilon_4 = 2t_y$, and $\epsilon_2=\epsilon_3=0$.
Therefore, when $t_y$ decreases, 
the conductance shows peaks corresponding 
to the resonance occurring in the one-dimensional 
system  $g_{\rm 1d}^0(\epsilon_n)$.

\begin{figure}
\vspace{-3cm}
\centerline{ \vbox{ \epsfxsize=150mm \epsfbox {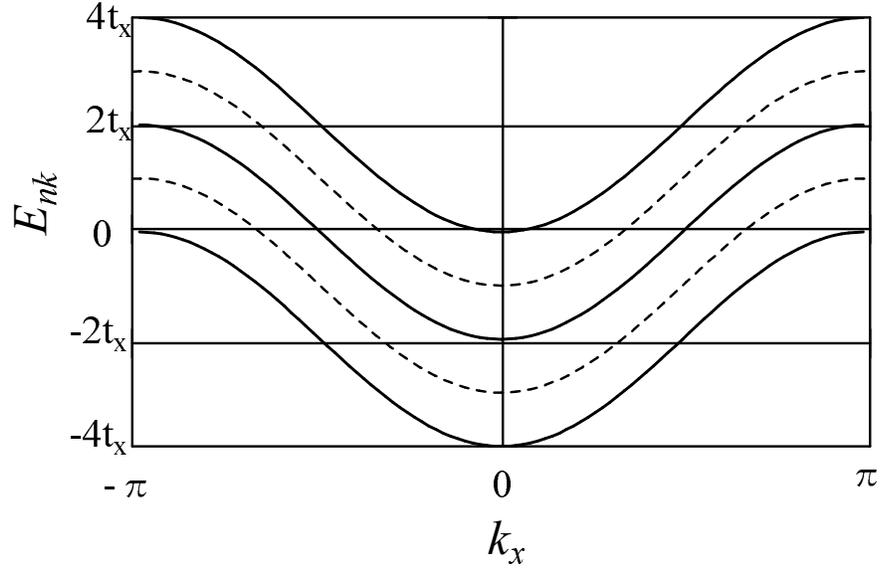}}
}
\vspace{-11cm}
\caption{Subband structure of the type I lead of the width $M=4$, 
where (solid line) $t_y/t_x=1.0$ and (dashed line) $t_y/t_x=0.5$. 
}
\label{fig:09}
\end{figure}

As an example of the the subband structure in the case of the type I leads,
we show the dispersion relation of an ideal system 
$E_{nk_x} = -2t_x \cos k_x + \epsilon_n$ in Fig.\ \ref{fig:09}.
Here the size along the $y$-direction is taken to be $M=4$, 
and the hopping matrix elements are taken to be 
 $t_y/t_x=1.0$ (solid line) and $t_y/t_x=0.5$ (dashed line). 
 In the case of $t_y/t_x =1.0$
 the highest and lowest subbands, i.e., modes $1$ and $4$, 
 do not contribute to the conductance 
 because the Fermi energy is situated at the band edge of 
 the subbands. 
 These two subband become  conducting for $0.0<t_y/t_x <1.0$. 
 We note that the modes $2$ and $3$ are degenerate and 
 several curves are overlaps on the line corresponding 
 to $-2t_x \cos k_x$ in Fig.\ \ref{fig:09}.

\subsection{Hubbard model connected to the type I leads}

\begin{figure}
\vspace{-3cm}
\hspace{0.5cm}
\centerline{ \vbox{ \epsfxsize=150mm \epsfbox {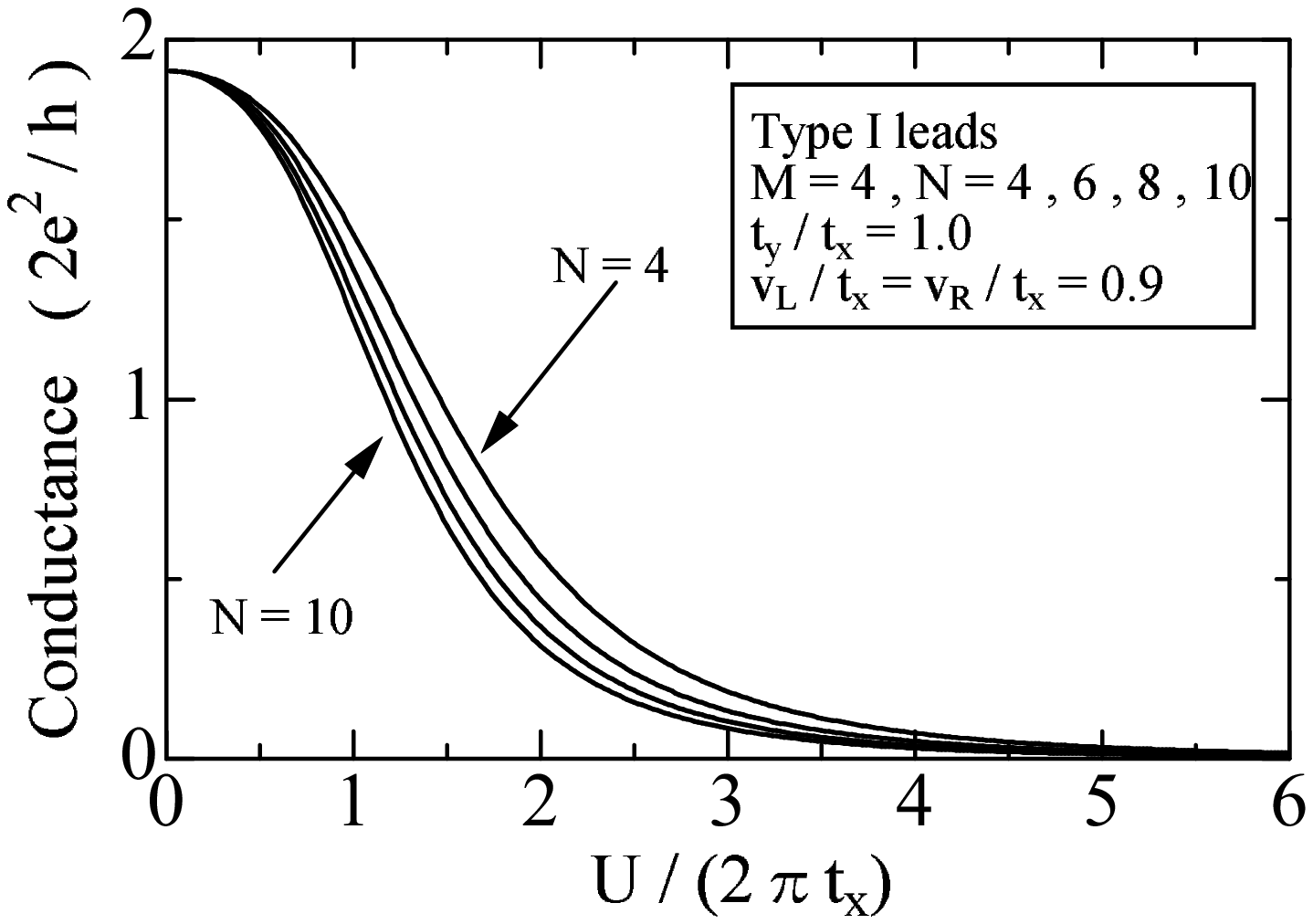}}
}
\vspace{-13cm}
\caption{
Conductance vs.\ $U$ in the isotropic case of $t_y/t_x=1.0$. 
The size of the interacting region is chosen to be 
$M=4$ and $N =4,6,8,10$. Type I leads. }
\label{fig:30}
\end{figure}

In this subsection, we discuss the results for
the Hubbard model connected to the type I leads. 
The conductance $g$ for 
the isotropic hopping $t_y/t_x=1.0$ is shown as a function of 
$U$ in Fig.\ \ref{fig:30}, 
where the size of the interacting region 
in the $y$-direction is $M=4$ and that 
in the $x$-direction is $N=4,6,8$, and $10$. 
The conductance decreases with increasing $U$.
The $N$ dependence is similar to that
in the one-dimensional case,
i.e.,  $g$ decreases monotonically with increasing $N$ showing a tendency
toward the Mott-Hubbard insulator.\cite{ao9} 
As mentioned in the above, 
the modes $1$ and $4$ are not conducting in the case of $t_y/t_x=1.0$,  
and the conductance is determined by 
the contributions of the modes $2$ and $3$.
This situation is changed in the anisotropic cases  $t_y/t_x <1.0$.

\begin{figure}
\vspace{-3cm}
\centerline{ \vbox{ \epsfxsize=150mm \epsfbox {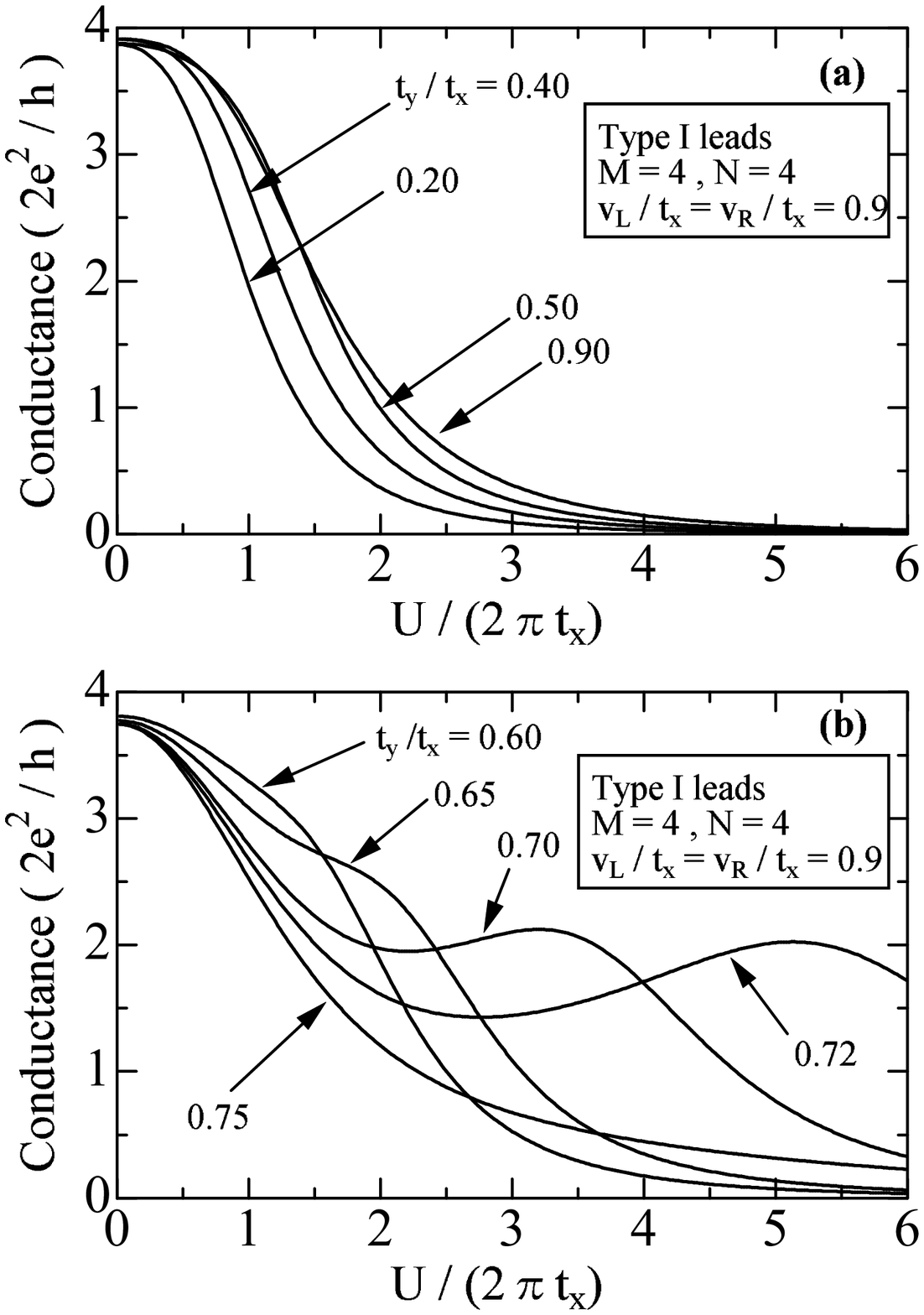}}
}
\vspace{-5cm}
\caption{
Conductance vs.\ $U$ in the anisotropic case $t_y/t_x<1.0$.
The size of the interacting region is chosen to be 
$M=4$ and $N =4$. Type I leads. 
}
\label{fig:31}
\end{figure}

\begin{figure}
\vspace{-3cm}
\centerline{ \vbox{ \epsfxsize=150mm \epsfbox {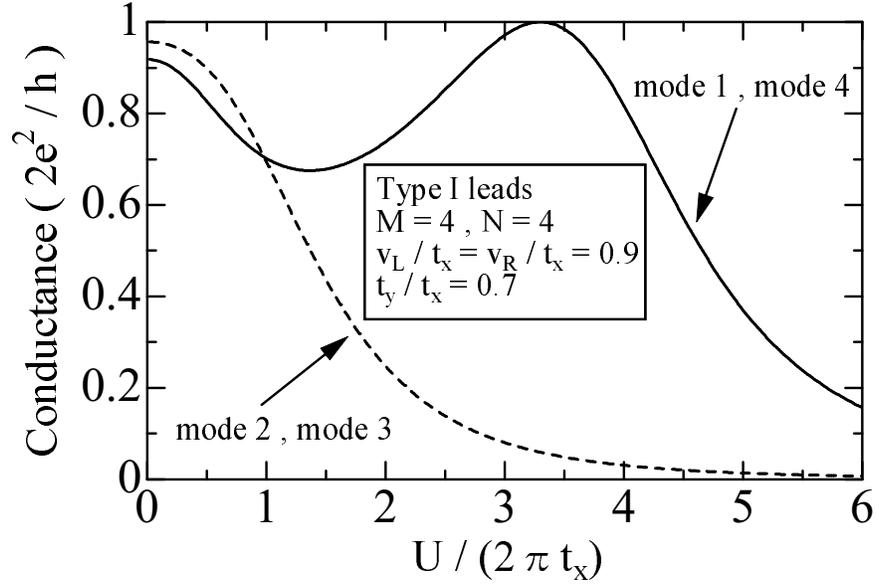}}
}
\vspace{-13cm}
\caption{
Conductance of each subband, 
where $M=4$, $N=4$, and $t_y/t_x=0.7$. Type I leads.
}
\label{fig:35}
\end{figure}

\begin{figure}
\vspace{-1cm}
\hspace{0.5cm}
\centerline{ \vbox{ \epsfxsize=150mm \epsfbox {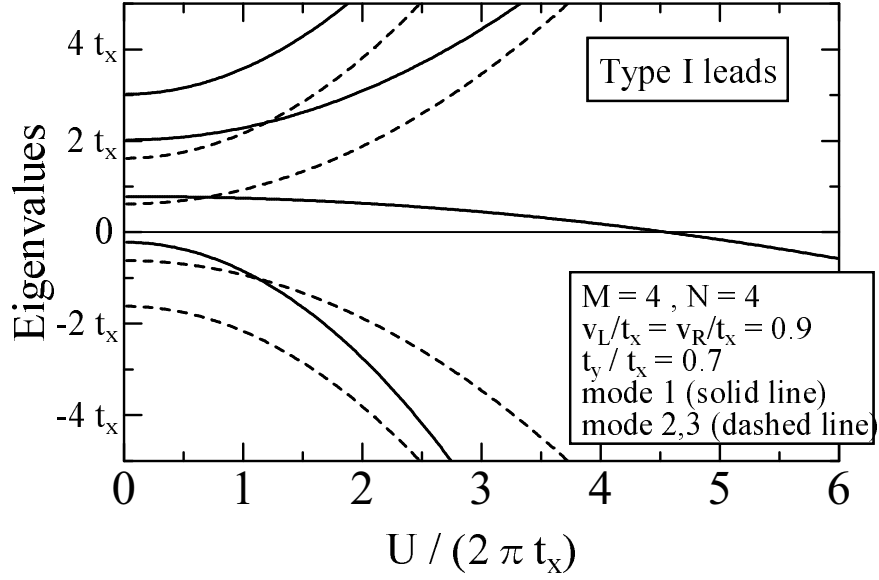}}
}
\vspace{-13cm}
\caption{Eigenvalues of 
$\widetilde{\mbox{\boldmath ${\cal H}$}}_{\rm C}^{(n)}$ 
for $n=1$ (solid line)  and $n=2,3$ (dashed line),
where $M=4$,  $N=4$, and $t_y/t_x=0.7$. 
The origin of the energy is set to be $E_F=0$.
Type I leads. 
}
\label{fig:50}
\end{figure}

In Fig.\ \ref{fig:31}, 
the conductance is plotted for several values of $t_y/t_x$, where 
the size of the interacting region is taken to be $N=4$ and $M=4$.
There is a broad peak at finite $U$ 
for $0.65 \lesssim t_y/t_x \lesssim 0.75$.
This shows that there are parameter regions where
the total conductance increases 
with $U$ even in the half-filled case. 
To clarify this behavior,
in Fig.\ \ref{fig:35} the contributions of each of the conducting modes
are plotted separately taking $t_y/t_x$ to be $0.7$. 
The resonant tunneling occurs in the modes $1$ and $4$ 
at $U/(2 \pi t_x) \simeq 3.3$,
while the contributions of the modes $2$ and $3$ decreases monotonically 
with increasing $U$.
Note that the pair of the subbands 
whose wavenumber in the $y$-direction are $k_y$ and $-k_y$ 
give the same contributions to the conductance.
Furthermore in the electric-hole symmetric case,
the contributions of the subbands whose eigenvalues 
are $\epsilon_n$ and $-\epsilon_n$ 
are the same.
The occurrence of the resonant tunneling  
is linked with the behavior of the eigenvalues 
of effective Hamiltonian 
which is defined in terms of the renormalized hopping matrix element 
$-\widetilde{t}^{(n)}_{ll'}$ given by eq.\ (\ref{eq:25}). 
In Fig.\ \ref{fig:50}, the eigenvalue 
of the $N\times N$ matrix $\widetilde{\mbox{\boldmath ${\cal H}$}}_{\rm C}^{(n)}$ 
is plotted as a function of $U$ for the mode $1$ (solid line), 
where $t_y/t_x=0.7$, $N=4$ and $M=4$.
The second lowest eigenvalue crosses the Fermi energy 
at $U/(2 \pi t_x) \simeq 4.6$. This corresponds to
the peak seen at $U/(2 \pi t_x) \simeq 3.3$ in Fig.\ \ref{fig:35}, 
although the values of $U$ do not coincide. 
This difference is mainly due to  
the contribution of the real part 
of the mixing term eq.\ (\ref{eq:87}):  
$\mbox{Re} \widehat{\mbox{\boldmath ${\cal V}$}}_{\rm mix}^+(0)$ 
causes the energy shift,
and the position of the resonance corresponding 
to the second eigenvalue moves toward the low-energy side.
While the transmission probability is defined with respect to 
the effective Hamiltonian of the whole system
 $\widetilde{\cal H}_{\rm qp}$ in eq.\ (\ref{eq:H_eff}),
the eigenvalues of the partial Hamiltonian 
 $\widetilde{\mbox{\boldmath ${\cal H}$}}_{\rm C}^{(n)}$ 
are useful to investigate the behaviors of the conductance.
Note that $\widetilde{t}^{(n)}_{ll'}$ depends on 
the mixing matrix elements 
$v_{\rm L}^{\phantom{0}}$ and $v_{\rm R}^{\phantom{0}}$ 
because the unperturbed Green's functions used to 
calculate the self-energy eq.\ (\ref{eq:Self_2v}) are defined 
with respect to the whole system. 
In Fig.\ \ref{fig:50}, the eigenvalues for the modes $2$ and $3$ are also 
plotted (dashed lines). 
In the present case, 
the eigenvalues of the modes $2$ and $3$ are the same,
and the eigenvalues of the modes $1$ and $4$ are symmetric 
with respect to the Fermi energy $\omega=0$.
The energy gap between the second and 
third eigenvalues of the mode $2$ (or $3$) 
becomes large with increasing $U$,
and it seems to show the tendency toward the Mott-Hubbard insulator.

\begin{figure}
\vspace{-3cm}
\centerline{ \vbox{ \epsfxsize=150mm \epsfbox {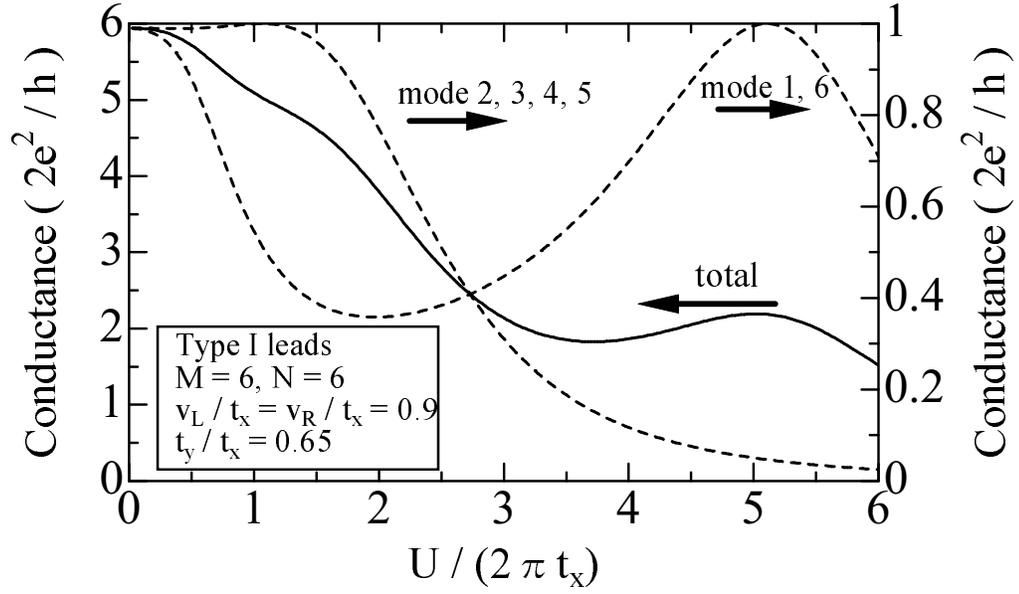}}
}
\vspace{-13cm}
\caption{
Conductance vs.\ $U$ for the system of the size $M=6$ and $N=6$: 
 (solid line) total conductance
and 
(dashed line) contributions of each mode for $t_y/t_x=0.65$. 
Type I leads.
}
\label{fig:57}
\end{figure}

We have also examined the conductance in the case of $M=6$,
where the subbands are classified into two groups;
$\epsilon_1=-\epsilon_6=2t_y$ and 
$\epsilon_2=\epsilon_3=-\epsilon_4=-\epsilon_5=t_y$.
In Fig.\ \ref{fig:57}, the total conductance (solid line) and
the contributions of the two groups of 
subbands (dashed line) are shown for $t_y/t_x=0.65$, 
where the size of the interacting region is taken to be $N=6$ and $M=6$. 
The total conductance show two bump like behaviors 
at $U/(2\pi t_x)\simeq 1.5$ and $5.1$,
which correspond to the resonance occurring in each group of the subbands. 
The resonance occurs at different values of $U$ for 
the different groups of the subbands.
This seems to show a general tendency that
the resonance does not occur simultaneously 
in the different groups of the subbands.
Thus, when the resonance occurs at one group, 
the remaining $M-2$ or $M-4$ subbands behave monotonically. 
Therefore the presence of the visible maximum in the $U$ dependence of
the total conductance is expected only for small $M$ where 
the contributions of the resonating subbands are comparable to 
the those of the remaining subbands.

\begin{figure}
\vspace{-2cm}
\centerline{ \vbox{ \epsfxsize=150mm \epsfbox {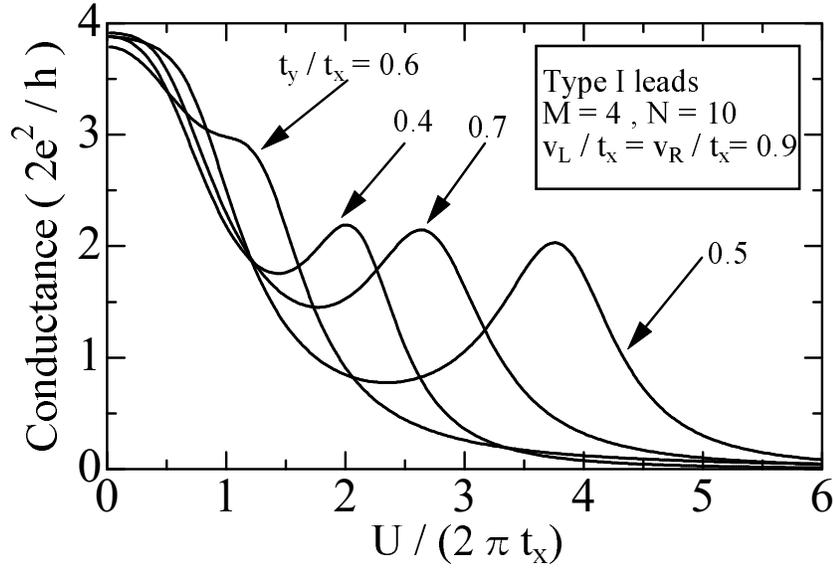}}
}
\vspace{-12cm}
\caption{
Conductance vs.\ $U$ for the system of the size $M=4$ and $N=10$, where  
$t_y/t_x = 0.4, 0.5, 0.6,$ and $0.7$. Type I leads.
}
\label{fig:24}
\end{figure}

\begin{figure}
\hspace{1cm}
\centerline{ \vbox{ \epsfxsize=150mm \epsfbox {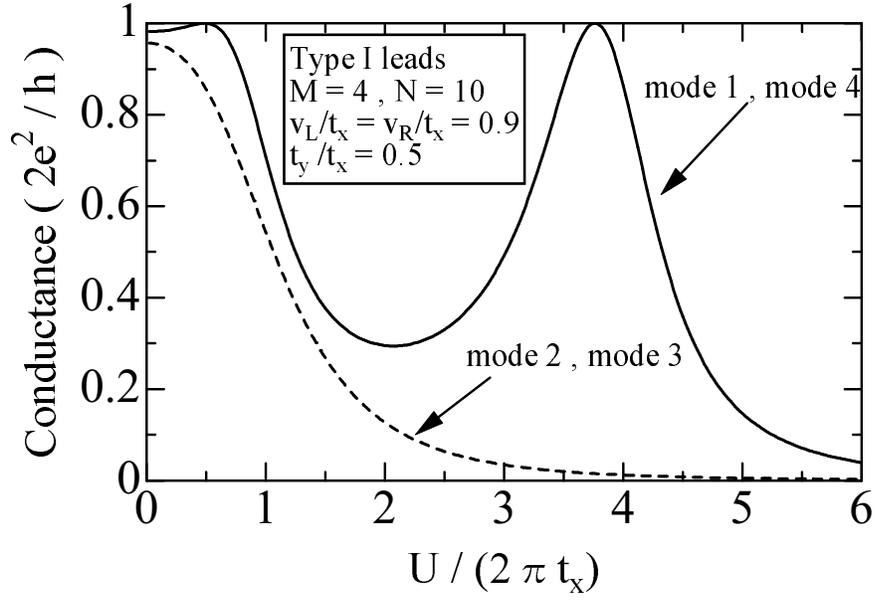}}
}
\vspace{-10cm}
\caption{
Conductance of each subband for the system of the size $M=4$ and $N=10$,
where  $t_y/t_x=0.5$. Type I lead. 
}
\label{fig:51}
\end{figure}

\begin{figure}
\vspace{-4cm}
\centerline{ \vbox{ \epsfxsize=150mm \epsfbox {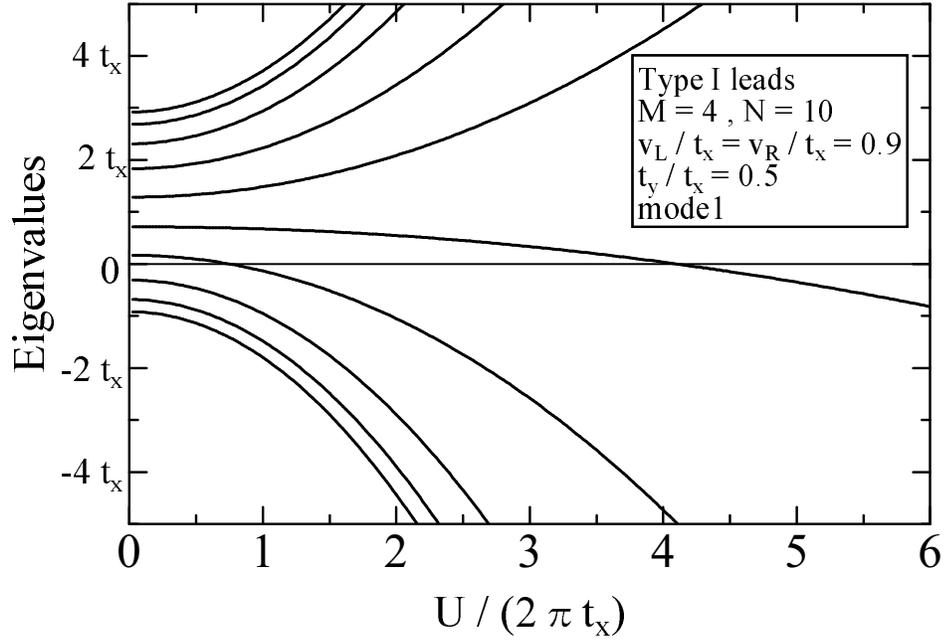}}
}
\vspace{-10cm}
\caption{Eigenvalues of 
$\widetilde{\mbox{\boldmath ${\cal H}$}}_{\rm C}^{(n)}$ for $n=1$, 
where $M=4$, $N=10$, and $t_y/t_x=0.5$. 
The origin of the energy is set to be $E_F=0$. Type I leads.
}
\label{fig:52}
\end{figure}

The number of the eigenvalues which cross the
Fermi energy is expected to increase 
with the size in the $x$-direction $N$. 
To clarify this feature, we calculate the conductance 
for $N=10$ keeping the width $M=4$ unchanged.
The results are shown in Fig.\ \ref{fig:24} for several values of $t_y/t_x$.
In each of the cases, at least one peak is visible at finite $U$.
Specifically in the case of $t_y/t_x=0.5$,  a shoulder like structure 
is seen at $U/2\pi t_x \simeq 0.5$ in addition 
to the peak at $U/2\pi t_x \simeq 3.8$.
To see this behavior precisely, 
the contributions of each of the subbands 
are plotted in Fig.\ \ref{fig:51}.
Two resonant peaks are seen
at $U/(2 \pi t_x) \simeq 0.5$ and $3.8$ in the modes $1$ and $4$.
The $U$ dependence of the eigenvalues of 
 $\widetilde{\mbox{\boldmath ${\cal H}$}}_{\rm C}^{(n)}$ 
for $n=1$ (mode $1$) is shown in Fig.\ \ref{fig:52}.
The 4th and 5th lowest eigenvalues cross the Fermi energy
at $U/(2 \pi t_x) \simeq 0.7$ and $4.1$, respectively.
These correspond to the two resonant peaks seen in Fig.\ \ref{fig:51}
as in the case of $N=4$.
The energy shift caused by the mixing with the leads
seems to decrease with increasing $N$.

\subsection{Hubbard model connected to the type II leads}
\begin{figure}
\vspace{-3cm}
\centerline{ \vbox{ \epsfxsize=150mm \epsfbox {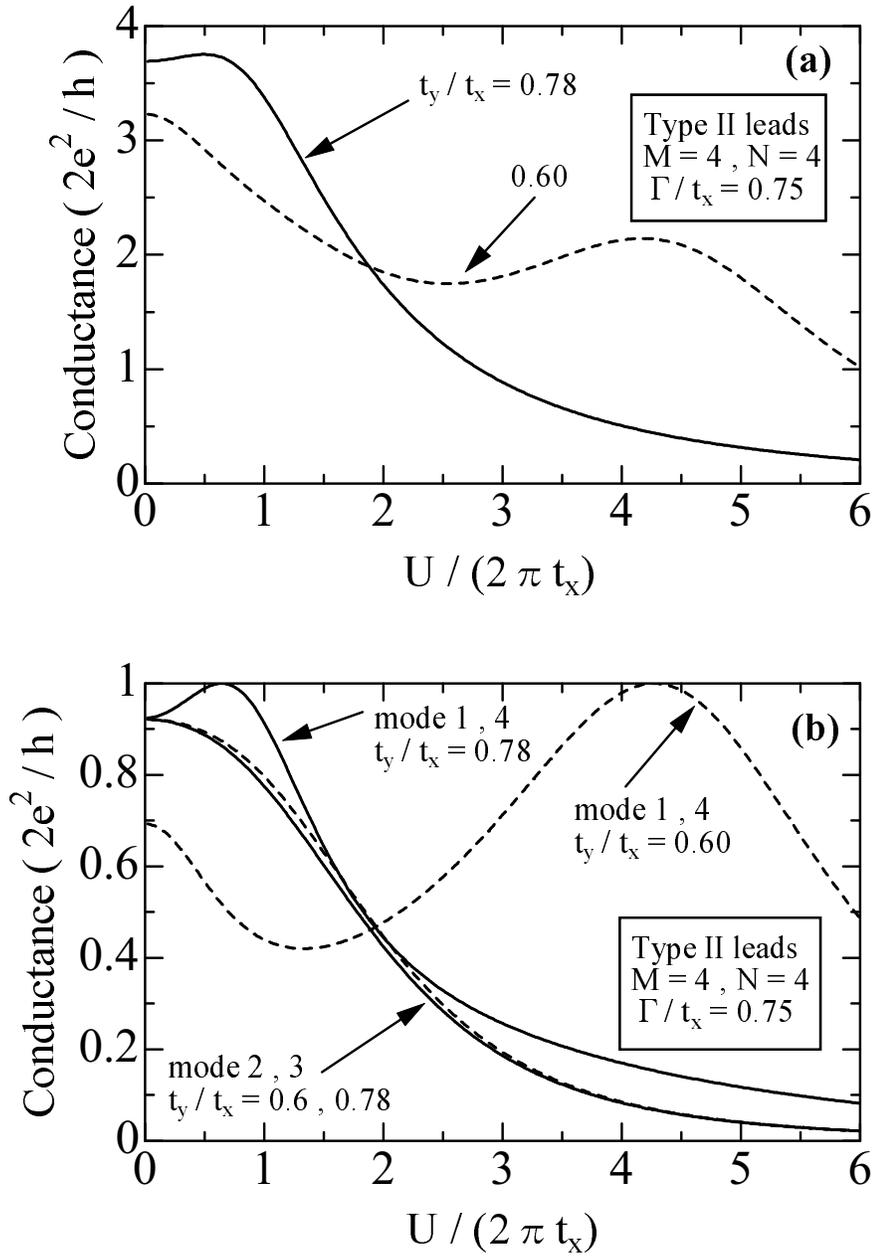}}
}
\vspace{-4cm}
\caption{
Conductance vs.\ $U$ for the system of 
$M=4$ and $N=4$ in the case of the type II leads: 
(a) total conductance, and 
(b) contributions of each mode. 
The hopping matrix element is taken to be 
 (dashed line)  $t_y/t_x=0.6$, and (solid line) $0.78$.
}
\label{fig:54}
\end{figure}

\begin{figure}
\vspace{-3cm}
\centerline{ \vbox{ \epsfxsize=150mm \epsfbox {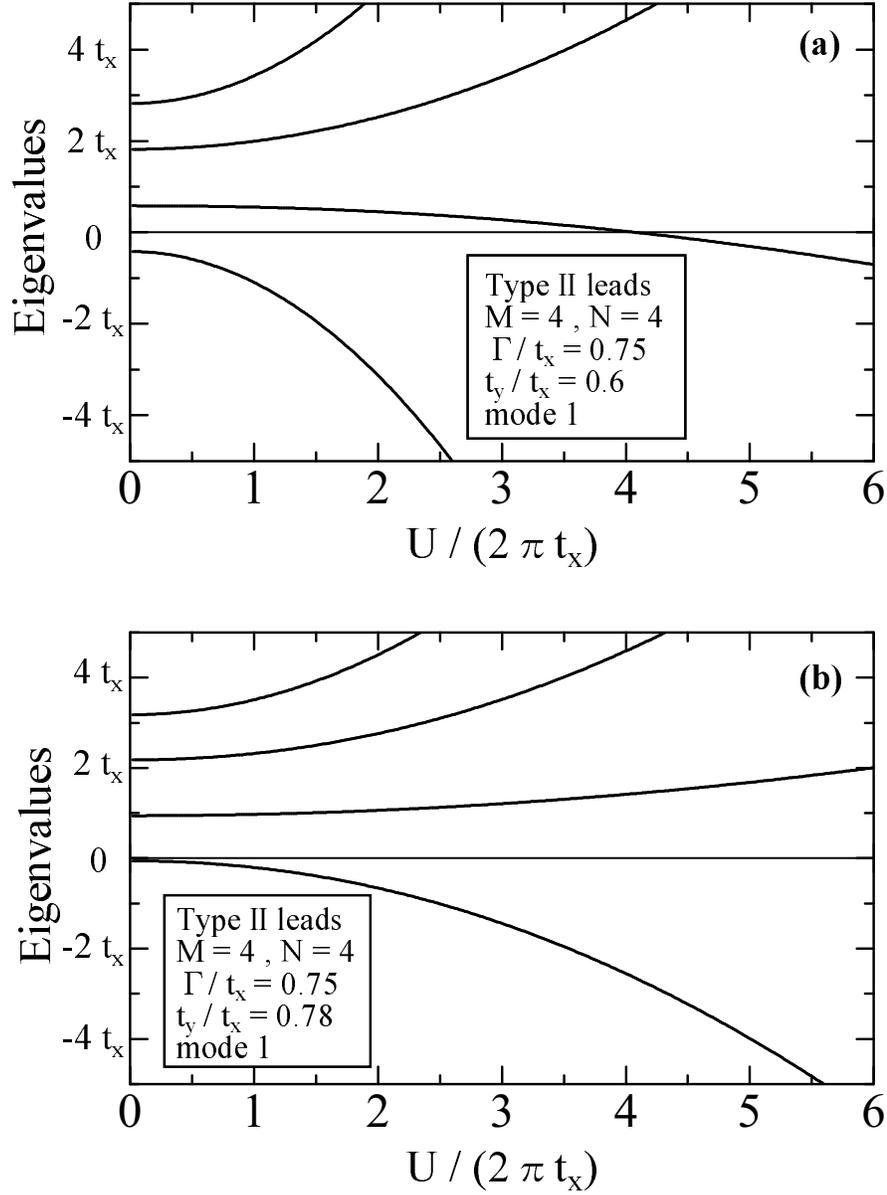}}
}
\vspace{-5cm}
\caption{Eigenvalues of 
$\widetilde{\mbox{\boldmath ${\cal H}$}}_{\rm C}^{(n)}$ for $n=1$, 
where $M=4$ and $N=4$. The hopping matrix element is 
taken to be  (a) $t_y/t_x=0.6$ and (b) $0.78$.
The origin of the energy is set to be $E_F=0$.
Type II leads.
}
\label{fig:55}
\end{figure}

We next discuss the conductance of 
the Hubbard model connected to the type II leads. 
In this lead, the local density of states at the interface 
of the lead is a constant and the bandwidth is infinity.
Then $\widehat{\mbox{\boldmath ${\cal V}$}}_{\rm mix}^+(\omega)$ 
becomes pure imaginary and independent of $\omega$, 
i.e., 
$\mbox{Re}\, \widehat{\mbox{\boldmath ${\cal V}$}}_{\rm mix}^+(\omega)
 \equiv 0$  and 
$\partial \widehat{\mbox{\boldmath ${\cal V}$}}_{\rm mix}^+(\omega)
/\,\partial \omega  \equiv 0$.  
Therefore the energy shift caused by the mixing with the leads
is expected to be small compared to that in the case of the leads I. 
In Fig.\ \ref{fig:54}\ (a),
the conductance in the case of $t_y/t_x=0.6$ and $0.78$ is shown 
as a function of $U$,
where the size of the interacting region is chosen to be $N=4$ and $M=4$.
There is a resonance peak at finite $U$ in each line; 
at $U/(2 \pi t_x) \simeq 4.3$ for $t_y/t_x=0.6$ 
and  $U/(2 \pi t_x) \simeq 0.6$ for $t_y/t_x=0.78$. 
The contributions of each of the subbands are plotted separately
in Fig.\ \ref{fig:54}\ (b). 
The peak seen in the total conductance is due to the resonance
tunneling in the modes $1$ and $4$.
Qualitatively, these results are similar to those of the type I leads.
In Fig.\ \ref{fig:55}, 
the eigenvalues of the effective 
Hamiltonian $\widetilde{\mbox{\boldmath ${\cal H}$}}_{\rm C}^{(n)}$ for 
$n=1$ (mode $1$) are plotted 
for (a) $t_y/t_x=0.6$ and (b) $t_y/t_x=0.78$. 
The second lowest eigenvalue 
for $t_y/t_x=0.6$ becomes zero at $U/(2 \pi t_x) \simeq 4.1$. 
This is close to the position of the resonance peak 
seen in Fig.\ \ref{fig:54}\ (b). 
The lowest eigenvalue  for $t_y/t_x=0.78$ 
does not cross the Fermi energy for any values of $U$,
although it is nearly zero for small $U$.
Nevertheless, this eigenstate corresponds to the resonance peak seen 
at $U/(2 \pi t_x) \simeq 0.6$ in Fig.\ \ref{fig:54}\ (b). 
This is because the lowest eigenvalue 
is shifted slightly up due to the mixing,
and it becomes to cross the Fermi energy.

\begin{figure}
\vspace{-3cm}
\centerline{ \vbox{ \epsfxsize=150mm \epsfbox {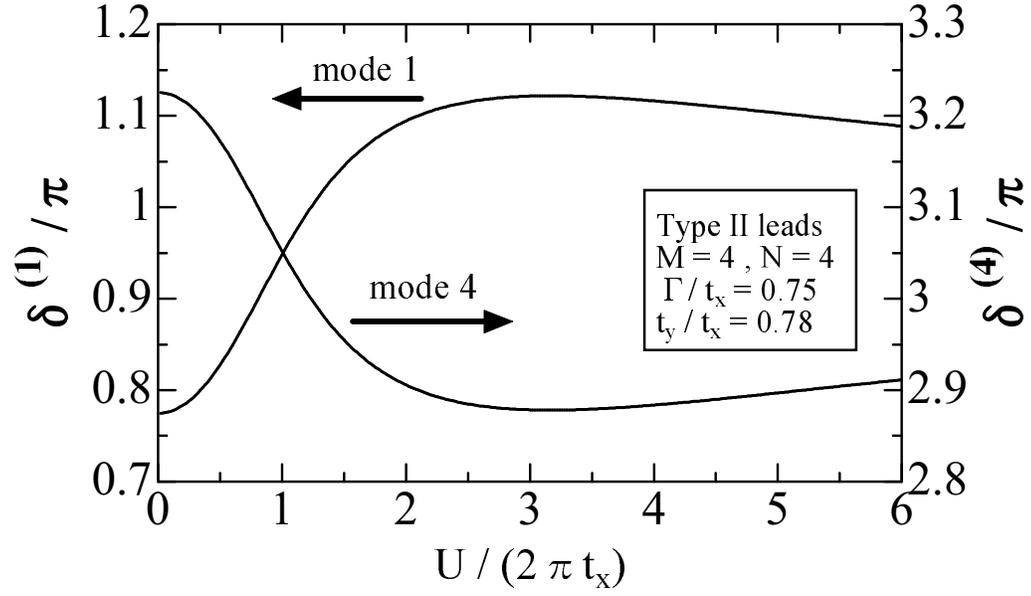}}
}
\vspace{-12cm}
\caption{
Phase shift 
$\delta^{(n)}/\pi$ vs. $U$ 
for $n=1$ and $n=4$, where $M=4$, $N=4$,
and $t_y/t_x=0.78$. Type II leads. 
}
\label{fig:56}
\end{figure}

We next discuss the phase shift $\delta^{(n)}$. 
As mentioned, the charge displacement of the $n$-th subband 
can be deduced from the phase shift using the Friedel sum rule 
$\Delta N_{\rm tot}^{(n)} = 2\, \delta^{(n)}/\pi$, 
where the factor $2$ corresponds to the spin degeneracy. 
Specifically, in the case of the type II leads, 
the compensation theorem holds.\cite{Anderson}
Then the second and the third term in the right-hand of 
eq.\ (\ref{eq:dn_def}) vanish because 
$\widehat{\mbox{\boldmath ${\cal V}$}}_{\rm mix}^+(\omega)$  
is independent of $\omega$ (see also Appendix \ref{sec:Friedel}). 
Thus the number of electrons in the type II leads 
is unchanged when the leads are connected to the interacting region, 
and the left-hand side of eq.\ (\ref{eq:Friedel_mode}),
$\Delta N_{\rm tot}$, 
becomes equal to the number of electrons in the interacting region 
$\sum_{i\in {\rm C},\sigma} 
 \langle c^{\dagger}_{i \sigma} c^{\phantom{\dagger}}_{i \sigma}
\rangle$.   
In Fig.\ \ref{fig:56}, the phase shift $\delta^{(n)}/\pi$ is 
plotted as a function of $U$ for the modes $1$ and $4$,  
where the parameters are taken to be 
$N=4$, $M=4$, and $t_y/t_x=0.78$. 
Here the principal value of the phase shift is determined 
in the limit of $U \to 0$ by comparing with the charge 
displacement obtained independently.
Reflecting the resonance seen in Fig.\ \ref{fig:54} (b), 
$\delta^{(n)}$ for the mode $1$ 
starts to increase rapidly at $U/(2 \pi t_x) \simeq 0.6$.
Correspondingly in the mode $4$, 
which is the lowest subband in this case, the phase shift 
$\delta^{(4)}$ decreases.
These behaviors show how the filling of the subband changes
when the resonance peak crosses the Fermi energy.
In Fig.\ \ref{fig:56}, 
the sum of the two phase shifts is a constant, i.e., 
$2\,\delta^{(1)}/\pi + 2\,\delta^{(4)}/\pi = 2\,N$ with $N=4$. 
The phase shifts for the modes $2$ and $3$ are the same, 
and it is a constant independent of $U$, i.e., 
$2\,\delta^{(n)}/\pi= N$ for $n=2$ and $3$ with $N=4$. 
This means that these two subbands are half filled.
Note that the total charge displacement is given by 
$\Delta N_{\rm tot}=M\times N$ in the electron-hole symmetric case.

\subsection{Remarks}
The results presented in this section 
are obtained using the order $U^2$ self-energy given 
by eq.\ (\ref{eq:Self_2v}).
Therefore, the results plotted for rather large 
values of $U$ may not be sufficient in quantitative sense. 
Nevertheless, some typical features of the results 
are seen even for small $U$.
For instance, the resonance at finite $U$ occurs 
for small value of $U/(2 \pi t_x) \lesssim 1$ in the case 
of $t_y/t_x=0.78$ as shown in Fig.\ \ref{fig:54}\ (a). 
Therefore we believe that the qualitative features of the results, 
such as the relation between the resonance and the
eigenvalues of the effective Hamiltonian, hold also for large $U$.  
The perturbation approach 
gives us a correct, in principle, description of 
the conductance and the total charge displacement 
with respect to the local Fermi-liquid ground state.

\section{Summary}
We have studied the conductance through  small interacting systems connected 
to multi-mode leads based on a local Fermi-liquid approach.
At $T=0$, the conductance and the total charge displacement are
determined by the value of the Green's function at the Fermi energy. 
Since the excitations at the Fermi energy do not decay at $T=0$,
there exists the one-particle Hamiltonian 
which reproduces these two quantities exactly. 
We have applied this formulation to a two-dimensional 
Hubbard model of finite size in the electron-hole symmetric case. 
We have calculated all the matrix elements of 
the order $U^2$ self-energy and determine 
the parameters of the effective Hamiltonian up to terms of order $U^2$.  
Specifically, we have examined two types of the noninteracting leads: 
I) semi-infinite tight-binding leads,
and II) leads of a constant density of states.
The results are similar, qualitatively, in both types of the leads. 
The conductance shows maximums in the $U$ dependence 
for some ranges of $t_y/t_x$,
where $t_x$ and $t_y$ are the hopping matrix element 
 in the $x$- and $y$-directions, respectively.
This means that there exists parameter regions, 
where the total conductance increases with $U$, even in the half-filled case.
By decomposing the total conductance into the sum of
the contributions of the subbands,
it is clarified that the peaks of the conductance are
caused by the resonance occurring in some group of the subbands 
which have the similar symmetric properties.
The phase shift of the subbands obtained from 
the Friedel sum rule shows a typical changes when the resonance occurs.
The resonance generally does not occur simultaneously 
in different groups of the subbands,  
and the subbands of off-resonance behave monotonically.
Therefore the maximum in the $U$ dependence of
the total conductance is expected only for the mesoscopic systems 
in which the number of the conducting modes is small enough 
and the contributions of the subbands of on-resonance are 
comparable to the those of the remaining subbands.

The perturbation approach examined in this work 
can be applied to interacting electrons in disordered systems. 
Especially, the analysis of the eigen values of the effective Hamiltonian,
which is used in the present study 
to investigate the behaviors of the resonance, 
may be combined with the Thouless-number\cite{LicciThouless} and 
finite-size scaling\cite{MackinnonKramer} methods.
The application along this line seems to be interesting 
in relation to the metal-insulator transition observed in 
two-dimensional systems.\cite{Kravchenko}
Furthermore, extensions to the finite temperatures\cite{ao11} and  
nonequilibrium situations\cite{ao10} are left for future studies.

\section*{Acknowledgements}
We would like to thank K.\ Murata, K.\ Tanigaki, and 
S.\ Nonoyama for valuable discussions.
Numerical computation was partly performed 
at computation center of Nagoya University and 
at Yukawa Institute Computer Facility.
This work was supported by the Grant-in-Aid 
for Scientific Research from the Ministry of Education, 
Science and Culture, Japan.

\appendix
\section{Conductance}
\label{sec:CONDUCTANCE}

In this appendix,
we provide the derivation of 
the dc conductance in the multi-mode case eq.\ (\ref{eq:g_multi}) 
by generalizing the derivation given in the single-mode case.\cite{ao6} 
The current operator corresponding 
to the mixing term eq.\ (\ref{eq:H_mix_multi}) is given by  
\begin{eqnarray}
 J_{\rm L}  &=& {\rm i}\, e 
         \sum_{m=1}^{M_{\rm L}} \sum_{\sigma}
                  v_{{\rm L},m}^{\phantom{\dagger}} 
           \left(\, 
            c^{\dagger}_{{\cal I}_m \sigma}\, 
            c^{\phantom{\dagger}}_{{\cal L}_m \sigma}  
          - c^{\dagger}_{{\cal L}_m \sigma}\, 
            c^{\phantom{\dagger}}_{{\cal I}_m\sigma}
\, \right) 
\label{eq:J_L}
\;,\\
 J_{\rm R}  &=& {\rm i}\, e 
        \sum_{m=1}^{M_{\rm R}} \sum_{\sigma}
         v_{{\rm R},m}^{\phantom{\dagger}}
  \left(\, 
           c^{\dagger}_{{\cal R}_m \sigma}\, 
           c^{\phantom{\dagger}}_{{\cal N}_m  \sigma} 
        -  c^{\dagger}_{{\cal N}_m \sigma}\, 
           c^{\phantom{\dagger}}_{{\cal R}_m  \sigma} 
\, \right) 
\label{eq:J_R} \;.
\end{eqnarray}
Here $J_{\rm L}$ is the total current flowing into the sample 
from the left lead, and 
$J_{\rm R}$ is the current flowing out to the right lead from the sample. 
These currents and the total charge in the sample 
$\rho_{\rm C}^{\phantom{0}} =  e \sum_{j\in {\rm C}, \sigma} 
 c^{\dagger}_{j \sigma} c^{\phantom{\dagger}}_{j \sigma}$
satisfy the equation of continuity  
$\partial \rho_{\rm C}^{\phantom{0}} 
/ \partial t \,+ J_{\rm R} - J_{\rm L} = 0$.
In the linear response theory, the dc conductance is given by
\begin{eqnarray}
 g \  &=& \  \lim_{\omega \to 0}
     { K_{\alpha\alpha'}^+(\omega) 
      - K_{\alpha\alpha'}^+(0) \over {\rm i} \omega } 
\label{eq:Kubo_M} \;, 
\\
K_{\alpha\alpha'}({\rm i} \nu_l) &=&   \int_0^{\beta} d\tau \,
\langle\, T_{\tau}\, J_{\alpha}(\tau)\, J_{\alpha'}(0) \, \rangle 
      \, {\rm e}^{{\rm i}\, \nu_l \tau}  \;, 
\end{eqnarray}
where $\alpha, \alpha' = {\rm L}, {\rm R}$,
and $K_{\alpha\alpha'}^+(\omega)$ is the retarded function
which is obtained from 
the Matsubara function $K_{\alpha\alpha'}({\rm i} \nu_l)$ 
by the analytic continuation ${\rm i}\nu_l \to \omega + {\rm i}0^+$. 
The dc conductance eq.\ (\ref{eq:Kubo_M}) corresponds to
the $\omega$-linear imaginary part of $K_{\alpha\alpha'}^+(\omega)$, 
and it is independent of $\alpha$ and $\alpha'$ 
owing to the current conservation.\cite{Fisher,Lee}
Moreover,  $K_{\alpha'\alpha}(z)  =  K_{\alpha\alpha'}(z)$ 
because of the time reversal symmetry,
and thus the imaginary part of $K_{\alpha\alpha'}^+(\omega)$  
corresponds to the discontinuity of $K_{\alpha\alpha'}(z)$ 
along the real axis in the complex $z$-plane.
Therefore, the dc conductance is equal to the coefficient of 
the $\nu\, \mbox{sgn}\, \nu$ term 
of $K_{\alpha\alpha'}({\rm i}\nu)$,\cite{Shiba}  
where $\mbox{sgn}\, \nu$ is the sign function.
In what follows, we extract this singular term 
from $K_{\alpha\alpha'}({\rm i}\nu)$ taking 
the current operators to be $\alpha={\rm R}$ and $\alpha'={\rm L}$.

At $T=0$, 
$K_{\rm RL}({\rm i}\nu)$ is written, 
treating the Matsubara frequencies to be continuous variables, as
[see also Fig.\ \ref{fig:multi}]

\begin{full}
\begin{eqnarray}
K_{\rm RL}({\rm i} \nu)  &=& 
K_{\rm RL}^{(a)}({\rm i} \nu) \ + \ K_{\rm RL}^{(b)}({\rm i} \nu) \; , 
\label{eq:K_nu_M} \\
K_{\rm RL}^{(a)}({\rm i} \nu) &=& e^2 
\sum_{\sigma} 
\int_{-\infty}^{+\infty} \! {{\rm d}\varepsilon \over 2 \pi} 
\nonumber \\
& & \times 
\mbox{Tr} 
\Bigl[\  
 \mbox{\boldmath $v$}_{\rm R}^{\phantom{\dagger}}
 \, 
 \mbox{\boldmath $G$}_{{\cal R}{\cal L}}({\rm i}\varepsilon + i\nu)
 \,
 \mbox{\boldmath $v$}_{\rm L}^{\phantom{\dagger}} 
  \, 
 \mbox{\boldmath $G$}_{{\cal I}{\cal N}}({\rm i}\varepsilon)
 +  
 \mbox{\boldmath $v$}_{\rm R}^{\phantom{\dagger}}
 \, 
 \mbox{\boldmath $G$}_{{\cal N}{\cal I}}({\rm i}\varepsilon + {\rm i}\nu)
 \,
 \mbox{\boldmath $v$}_{\rm L}^{\phantom{\dagger}} 
  \, 
 \mbox{\boldmath $G$}_{{\cal L}{\cal R}}({\rm i}\varepsilon)
\nonumber \\
& & \quad 
- \mbox{\boldmath $v$}_{\rm R}^{\phantom{\dagger}}
  \, 
  \mbox{\boldmath $G$}_{{\cal R}{\cal I}}({\rm i}\varepsilon + {\rm i}\nu)
  \,
  \mbox{\boldmath $v$}_{\rm L}^{\phantom{\dagger}} 
  \, 
  \mbox{\boldmath $G$}_{{\cal L}{\cal N}}({\rm i}\varepsilon)
- \mbox{\boldmath $v$}_{\rm R}^{\phantom{\dagger}}
  \, 
  \mbox{\boldmath $G$}_{{\cal N}{\cal L}}({\rm i}\varepsilon + {\rm i}\nu)
  \,
  \mbox{\boldmath $v$}_{\rm L}^{\phantom{\dagger}} 
  \, 
  \mbox{\boldmath $G$}_{{\cal I}{\cal R}}({\rm i}\varepsilon)
\  \Bigr] \;,  
\label{eq:K_nu_a_M}
\\ 
K_{\rm RL}^{(b)}({\rm i} \nu) &=&  e^2 
\sum_{\sigma \sigma '} 
\sum_{ \{j\} \in {\rm C}}
\int_{-\infty}^{+\infty} 
\! {{\rm d}\varepsilon {\rm d}\varepsilon' \over (2 \pi)^2} \; 
\Gamma_{\sigma \sigma ' ;\, \sigma ' \sigma}^{j_1 j_2;\, j_3 j_4}
({\rm i} \varepsilon + {\rm i} \nu, {\rm i} \varepsilon'+ {\rm i} \nu ;
\, {\rm i} \varepsilon' , {\rm i} \varepsilon ) 
\nonumber \\
& & \times \Bigl\{  
  \mbox{\boldmath $G$}_{{\rm C}{\cal L}}({\rm i}\varepsilon + i\nu)
\, 
  \mbox{\boldmath $v$}_{\rm L}^{\phantom{\dagger}} 
\,  \mbox{\boldmath $G$}_{{\cal I}{\rm C}}({\rm i}\varepsilon)
\, 
- 
  \mbox{\boldmath $G$}_{{\rm C}{\cal I}}({\rm i}\varepsilon + {\rm i}\nu)
\,   \mbox{\boldmath $v$}_{\rm L}^{\phantom{\dagger}} 
\,  \mbox{\boldmath $G$}_{{\cal L}{\rm C}}({\rm i}\varepsilon)
\Bigr\}_{j_1j_4} 
\nonumber \\
& & \times \Bigl\{
  \mbox{\boldmath $G$}_{{\rm C}{\cal N}}({\rm i}\varepsilon')
\, 
  \mbox{\boldmath $v$}_{\rm R}^{\phantom{\dagger}} 
\, 
  \mbox{\boldmath $G$}_{{\cal R}{\rm C}}({\rm i}\varepsilon' + {\rm i}\nu)
-
  \mbox{\boldmath $G$}_{{\rm C}{\cal R}}({\rm i}\varepsilon')
\, 
  \mbox{\boldmath $v$}_{\rm R}^{\phantom{\dagger}} 
\,
  \mbox{\boldmath $G$}_{{\cal N}{\rm C}}({\rm i}\varepsilon' + {\rm i}\nu)
\Bigr\}_{j_3j_2} 
\;.
\label{eq:K_nu_b_M}
\end{eqnarray}
\end{full}

\begin{figure}
\vspace{-1cm}
\hspace{3cm}
\centerline{ \vbox{ \epsfxsize=100mm \epsfbox {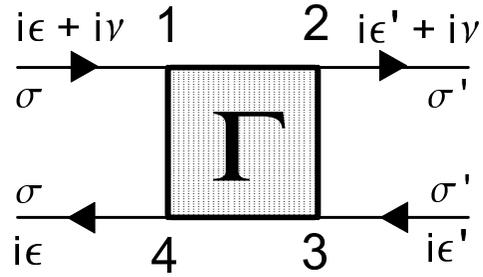}}
}
\vspace{-9cm}
\caption{The vertex function
$\Gamma_{\sigma \sigma ' ;\, \sigma ' \sigma}^{j_1 j_2;\, j_3 j_4} 
({\rm i} \varepsilon + {\rm i} \nu, {\rm i} \varepsilon'+ {\rm i} \nu ;
\, {\rm i} \varepsilon' , {\rm i} \varepsilon )$.
}
\label{fig:vertex}
\end{figure}

Here $\mbox{\boldmath $v$}_{\rm L}^{\phantom{\dagger}}$ 
and $\mbox{\boldmath $v$}_{\rm R}^{\phantom{\dagger}}$ are
diagonal matrices corresponding 
to $v_{{\rm L},m}^{\phantom{\dagger}}$ 
and $v_{{\rm R},m}^{\phantom{\dagger}}$, 
respectively. 
The subscript ${\rm C}$ denotes a set consisting 
of $N_{\rm C}$ sites in the central region.
$\mbox{\boldmath $G$}_{{\rm C}{\cal L}}(z)$ is a 
$N_{\rm C} \times M_{\rm L}$ matrix, and the $(j,m)$ element is given
by $G_{j,{\cal L}_m}(z)$ with $j \in {\rm C}$.  
Also, $\mbox{\boldmath $G$}_{{\cal R}{\rm C}}(z)$ is
a $M_{\rm R} \times N_{\rm C}$ matrix, and the $(m,j)$ element is given
by $G_{{\cal R}_m,j}(z)$ with $j \in {\rm C}$. 
The matrix Green's functions in 
eqs.\ (\ref{eq:K_nu_a_M}) and (\ref{eq:K_nu_b_M}) are defined in this way.
$\Gamma_{\sigma \sigma ' ;\, \sigma ' \sigma}^{j_1 j_2;\, j_3 j_4} 
({\rm i} \varepsilon_1, {\rm i} \varepsilon_2 
;\, {\rm i} \varepsilon_{3}, {\rm i} \varepsilon_{4} )$ 
is the vertex corrections due to the 
inter-electron interaction, and illustrated in Fig.\ \ref{fig:vertex}. 
Since we are considering the interaction which is switched on 
only in the central region, 
the Green's function satisfies following relations at the interfaces;  
\begin{eqnarray}
\left\{ 
\!\!\!
\begin{array}{ll}
  \mbox{\boldmath $G$}_{{\cal R} \gamma}(z) = 
           -
           \, \mbox{\boldmath $F$}_{\rm R}(z) 
           \,  \mbox{\boldmath $v$}_{\rm R}^{\phantom{\dagger}} 
           \, \mbox{\boldmath $G$}_{{\cal N}\gamma }(z) 
           &\mbox{for} \ \ 
               \gamma = {\cal L},\,{\cal I},\,{\rm C},\,{\cal N}  
           \\ 
  \mbox{\boldmath $G$}_{\gamma{\cal L}}(z) = 
           -
           \, \mbox{\boldmath $G$}_{\gamma{\cal I}}(z)  
           \, \mbox{\boldmath $v$}_{\rm L}^{\phantom{\dagger}} 
           \, \mbox{\boldmath $F$}_{\rm L}(z)
           &\mbox{for} \ \  
          \gamma = {\cal I},\,{\rm C},\,{\cal N},\, {\cal R}  
\end{array}  
\right. 
\nonumber \\
.
 \label{eq:rec_lead_M}
\end{eqnarray}
Using these relations, 
eqs.\ (\ref{eq:K_nu_a_M}) and (\ref{eq:K_nu_b_M}) are rewritten as 

\begin{full}
\begin{eqnarray}
K_{\rm RL}^{(a)}({\rm i} \nu) &=&  e^2 
\sum_{\sigma} 
\int_{-\infty}^{+\infty} \! {{\rm d}\varepsilon \over 2 \pi} 
\,
\mbox{Tr} \biggl[\, 
\mbox{\boldmath $v$}_{\rm R}^{\phantom{\dagger}}
\left[\,
 \mbox{\boldmath $F$}_{\rm R}({\rm i}\varepsilon + {\rm i}\nu)
 -
 \mbox{\boldmath $F$}_{\rm R}({\rm i}\varepsilon)
\,\right] 
 \, 
 \mbox{\boldmath $v$}_{\rm R}^{\phantom{\dagger}}
 \, 
 \mbox{\boldmath $G$}_{{\cal N}{\cal I}}({\rm i}\varepsilon + {\rm i}\nu)
 \nonumber \\
& &  \qquad \qquad \qquad \qquad \times
 \,
 \mbox{\boldmath $v$}_{\rm L}^{\phantom{\dagger}} 
  \, 
  \left[\,
 \mbox{\boldmath $F$}_{\rm L}({\rm i}\varepsilon + {\rm i}\nu)
 -
 \mbox{\boldmath $F$}_{\rm L}({\rm i}\varepsilon)
\,\right] 
 \,
 \mbox{\boldmath $v$}_{\rm L}^{\phantom{\dagger}} 
\, \mbox{\boldmath $G$}_{{\cal I}{\cal N}}({\rm i}\varepsilon)
\, \biggr]
 \;,
\label{eq:bubble_M} \\
K_{\rm RL}^{(b)}({\rm i} \nu) &=& e^2 
\sum_{\sigma \sigma '} 
\sum_{\{j\} \in {\rm C}}
\int_{-\infty}^{+\infty} 
\! {{\rm d}\varepsilon {\rm d}\varepsilon' \over (2 \pi)^2} \; 
\Gamma_{\sigma \sigma ' ;\, \sigma ' \sigma}^{j_1 j_2;\, j_3 j_4}
({\rm i} \varepsilon + {\rm i} \nu, {\rm i} \varepsilon'+ {\rm i} \nu 
;\, {\rm i} \varepsilon' , {\rm i} \varepsilon ) 
\nonumber \\
& & \times  
\Bigl\{ 
  \mbox{\boldmath $G$}_{{\rm C}{\cal I}}({\rm i}\varepsilon + {\rm i}\nu)
\, 
  \mbox{\boldmath $v$}_{\rm L}^{\phantom{\dagger}} 
  \, 
  \left[\,
 \mbox{\boldmath $F$}_{\rm L}({\rm i}\varepsilon + {\rm i}\nu)
 -
 \mbox{\boldmath $F$}_{\rm L}({\rm i}\varepsilon)
\,\right] 
\,   \mbox{\boldmath $v$}_{\rm L}^{\phantom{\dagger}} 
\,  \mbox{\boldmath $G$}_{{\cal I}{\rm C}}({\rm i}\varepsilon)
\Bigr\}_{j_1j_4} 
\nonumber \\
& & \times  \Bigl\{
  \mbox{\boldmath $G$}_{{\rm C}{\cal N}}({\rm i}\varepsilon')
\, 
  \mbox{\boldmath $v$}_{\rm R}^{\phantom{\dagger}} 
\, 
\left[\,
 \mbox{\boldmath $F$}_{\rm R}({\rm i}\varepsilon' + {\rm i}\nu)
 -
 \mbox{\boldmath $F$}_{\rm R}({\rm i}\varepsilon')
\,\right] 
\,
  \mbox{\boldmath $v$}_{\rm R}^{\phantom{\dagger}} 
\,
  \mbox{\boldmath $G$}_{{\cal N}{\rm C}}({\rm i}\varepsilon' + {\rm i}\nu)
\Bigr\}_{j_3j_2} 
\;.
\label{eq:vertex_M}
\end{eqnarray}
\end{full}

We can now extract the singular $\nu\, \mbox{sgn}\, \nu$ term
of $K_{\rm RL}({\rm i}\nu)$ using 
eqs.\ (\ref{eq:bubble_M}) and (\ref{eq:vertex_M}), 
as in the case of the single Anderson impurity.\cite{Shiba,ao3} 
It can be shown that there is no singular $\nu\, \mbox{sgn}\, \nu$ 
term in $K_{\rm RL}^{(b)}({\rm i}\nu)$ due the two factors 
$[ \mbox{\boldmath $F$}_{\rm L}({\rm i}\varepsilon + {\rm i}\nu) 
- \mbox{\boldmath $F$}_{\rm L}({\rm i}\varepsilon) ]$ 
and $[ \mbox{\boldmath $F$}_{\rm R}({\rm i}\varepsilon' + {\rm i}\nu) 
- \mbox{\boldmath $F$}_{\rm R}({\rm i}\varepsilon') ]$ which 
have different frequencies. 
Thus, the $\nu\, \mbox{sgn}\, \nu$ term comes 
only from $K_{\rm RL}^{(a)}({\rm i}\nu)$. It can be 
obtained from eq.\ (\ref{eq:bubble_M}) and yields eq.\ (\ref{eq:g_multi}).

\section{Friedel sum rule}
\label{sec:Friedel}

We provide here the outline of the derivation of
the Friedel sum rule in the single-mode case  
eq.\ (\ref{eq:Friedel}) following 
Langer and Ambegaokar.\cite{LangerAmbegaokar}
The Green's function in the left lead 
$ij \in {\rm L}$ is written as
\begin{eqnarray}
G_{ij}^{+}(\omega)  &=& F^{\rm L}_{ij}(\omega) \ 
+ \ F^{\rm L}_{i0}(\omega)\,  v_{\rm L}^{\phantom{\dagger}}
 \, G_{11}^{+}(\omega)\, v_{\rm L}^{\phantom{\dagger}}
 \, F^{\rm L}_{0j}(\omega) 
\;, 
\\
  F_{ij}^{\rm L}(\omega) &=&  
    \sum_n { \phi_{n,{\rm L}}^{\phantom{\dagger}}(i) \,\phi_{n,{\rm L}}^*(j)
              \over \omega - \epsilon_{n,{\rm L}} + {\rm i}0^+ } 
\;.
\end{eqnarray}
Here $F_{ij}^{\rm L}(\omega)$ is the retarded Green's function 
of the isolated lead, and 
$\epsilon_{n,{\rm L}}$ and $\phi_{n,{\rm L}}^{\phantom{\dagger}}(i)$ are
the eigenvalue and the corresponding eigenfunction 
of the one-particle Hamiltonian ${\cal H}_{\rm L}$.
Using the orthogonality relation 
$\,\sum_{i \in {\rm L}} \phi_{n,{\rm L}}^{\phantom{\dagger}}(i)\, 
\phi_{n',{\rm L}}^*(i) 
= \delta_{nn'}$, 
we obtain
\begin{eqnarray}
\sum_{i \in {\rm L}} \left[\, G_{ii}^{+} - F^{\rm L}_{ii}\,\right]
\ = \ - \, {\partial F^{\rm L}_{00} \over \partial \omega} 
\, v_{\rm L}^{\phantom{\dagger}} \, G_{11}^{+} 
\, v_{\rm L}^{\phantom{\dagger}}
\;.
\label{eq:dn_L}
\end{eqnarray}
Here $F^{\rm L}_{00} \equiv F_{\rm L}^{+}$, 
which is the Green's function at the interface.
Using eq.\ (\ref{eq:dn_L}) and 
the corresponding relation for the right lead,
the displacement of the total charge defined 
by eq.\ (\ref{eq:dn_def}) is written, at $T=0$, as

\begin{full}
\begin{eqnarray}
\Delta N_{\rm tot} 
&=&  - \, {2 \over \pi} \ \mbox{Im} \, \int_{-\infty}^0 {\rm d}\omega
\Biggl(\, \,  
\sum_{i \in {\rm C}} G_{ii}^+  
\ + \ \sum_{i \in {\rm L}} \left[\, G_{ii}^+ - F^{\rm L}_{ii}\,\right] 
\ + \ \sum_{i \in {\rm R}} \left[\, G_{ii}^+ - F^{\rm R}_{ii}\,\right]
\, \Biggr)
\nonumber\\
&=&  - \, {2 \over \pi} \ \mbox{Im} \, \int_{-\infty}^0 {\rm d}\omega
\ \mbox{Tr} \left[\, 
 \mbox{\boldmath ${\cal G}$}^+ 
  \, - \,   \mbox{\boldmath ${\cal G}$}^+ \, 
{\partial \mbox{\boldmath ${\cal V}$}_{\rm mix}^+
\over \partial \omega}
    \,\right]
\nonumber\\
&=&  
  - \, {2 \over \pi} \ \mbox{Im} \, \int_{-\infty}^0 {\rm d}\omega  
\left(\, 
{\partial \over \partial \omega} \ \mbox{Tr} \left[\, 
 \log \left\{ \mbox{\boldmath ${\cal G}$}^+ \right\}^{-1} \,\right] \  
 + \
 \mbox{Tr} \left[\,  \mbox{\boldmath ${\cal G}$}^+ \, 
{\partial  
        \mbox{\boldmath $\Sigma$}^+
\over   \partial \omega} 
\, \right] \, \right) .
\label{eq:dn}
\end{eqnarray}
\end{full}

Here we have used a matrix notation 
as that was used in the text eqs.\ (\ref{eq:81})-(\ref{eq:86}) 
for the multi-mode case:
the Green's function in the central region is denoted by
$\mbox{\boldmath ${\cal G}$}(z) = \{G_{ij}(z)\}$ with $ij \in  {\rm C}$
and the Dyson equation is given by 
$
\left\{\mbox{\boldmath ${\cal G}$}(z)\right\}^{-1}  
  =
z \, \mbox{\boldmath $1$} 
-  \mbox{\boldmath ${\cal H}$}_{\rm C}^0  
- \mbox{\boldmath ${\cal V}$}_{\rm mix}(z)  
- \mbox{\boldmath $\Sigma$}(z)$, 
where
$\mbox{\boldmath ${\cal H}$}_{\rm C}^0 =  
\{  -t_{ij}^{\rm C}-\mu \,\delta_{ij} \}$, and 
$\mbox{\boldmath ${\cal V}$}_{\rm mix}(z)$  
and 
$\mbox{\boldmath $\Sigma$}(z)$ are the single-mode version of
eqs.\ (\ref{eq:87}) and (\ref{eq:86}), respectively.

It has been shown that the contribution of the second term 
in the third line of eq.\ (\ref{eq:dn}) 
vanishes.\cite{LuttingerWard,Luttinger} 
Thus,  eq.\ (\ref{eq:dn}) can be rewritten as 
\begin{eqnarray}
\Delta N_{\rm tot} 
&=&
    {1\over  \pi {\rm i}}\, 
 \log \left[ 
 \det\left\{ \mbox{\boldmath ${\cal G}$}^-(0) \right\}^{-1} / 
 \det\left\{ \mbox{\boldmath ${\cal G}$}^+(0) \right\}^{-1} \right]
\label{eq:detS} \\
&=&
    {1\over  \pi {\rm i}}\, 
\log \, \det\left[ \mbox{\boldmath $1$}  
+ 2 {\rm i} \,\mbox{Im} \left\{ \mbox{\boldmath ${\cal V}$}_{\rm mix}^+(0) 
\right\} 
         \mbox{\boldmath ${\cal G}$}^+(0)
\right]  .
\label{eq:dn_det}
\end{eqnarray}
For obtaining eq.\ (\ref{eq:dn_det}),
we have used the Fermi-liquid property 
$\mbox{Im}  \mbox{\boldmath $\Sigma$}^+(0)$.
The matrix $\mbox{\boldmath ${\cal V}$}_{\rm mix}$ has only two
non-zero elements as eq.\ (\ref{eq:87}).
Therefore, the determinant in eq.\ (\ref{eq:dn_det}) can 
be written in terms of a $2\times 2$ matrix as  
\begin{eqnarray}
& &\Delta N_{\rm tot}
=
{1 \over \pi {\rm i}}\,  
\log [\, \det 
 \mbox{\boldmath $S$}
\,]  \;, 
\label{eq:Friedel} 
\\
\nonumber
\\
& & \mbox{\boldmath $S$}
 =  
  \left [
 \matrix { 1  & 0   \cr
           0  & 1  \cr  }
  \right ]  
- \, 2 {\rm i} 
  \left [ 
 \matrix { \Gamma_{\rm L}(0)  & 0            \cr
           0            & \Gamma_{\rm R}(0)  \cr  }
  \right ]  
  \left [ 
 \matrix { G_{1 1}^{+}(0)  & G_{1 N}^{+}(0)  \cr 
           G_{N 1}^{+}(0)  & G_{N N}^{+}(0)  \cr  }
  \right ] 
.   
\nonumber\\
\label{eq:S}
\end{eqnarray}

\clearpage

\clearpage

\clearpage

\end{document}